\documentclass[letterpaper,journal]{IEEEtran}

\usepackage{graphicx}
\graphicspath{{./Figures/}}
\usepackage{subfigure} 
\usepackage{xcolor,graphicx,float} 
\usepackage{amsthm}
\usepackage{amsmath}
\usepackage{multirow}
\usepackage{amssymb}
\usepackage{amsfonts}
\usepackage{algorithmic}
\usepackage{algorithm}
\usepackage{booktabs}
\usepackage{cite}
\usepackage[colorlinks,linkcolor=black,citecolor=black,urlcolor=black]{hyperref}
\usepackage{url}
\usepackage{titlesec}

\begin{document}
\title{Enhancing Cyber-Resilience in Integrated Energy System Scheduling with Demand Response Using Deep Reinforcement Learning}

\author{{Yang~Li, ~\IEEEmembership{Senior Member,~IEEE,} Wenjie~Ma, Yuanzheng~Li,~\IEEEmembership{Senior Member,~IEEE}, Sen Li, Zhe Chen,~\IEEEmembership{Fellow,~IEEE,} Mohammad Shahidehpour,~\IEEEmembership{Life Fellow,~IEEE}}

\thanks{ This work is supported by the National Natural Science Foundation of China under Grant (No. U2066208).}

\thanks{Y. Li is with the School of Electrical Engineering, Northeast Electric Power University, Jilin 132012, China (e-mail: liyang@neepu.edu.cn).}%

\thanks{W. Ma is with the State Grid XiXian New Area Power Supply Company, XiXian New Area 712000, China (e-mail: 2202100232@neepu.edu.cn).}%

\thanks{Y. Z. Li is with the School of Artificial Intelligence and Automation, Huazhong University of Science and Technology, Wuhan 430074, China (e-mail: Yuanzheng\_Li@hust.edu.cn).}

\thanks{S. Li is with the Department of Civil and Environmental Engineering, The Hong Kong University of Science and Technology, Hong Kong (e-mail: cesli@ust.hk).}

\thanks{Z. Chen is with the Institute of Energy Technology, Aalborg University, Aalborg DK-9220, Denmark (e-mail: zch@energy.aau.dk).}

\thanks{M. Shahidehpour is with the ECE Department, Illinois Institute of Technology, Chicago, IL 60616, USA (e-mail: ms@iit.edu).}}

\maketitle

\begin{abstract}
Optimally scheduling multi-energy flow is an effective method to utilize renewable energy sources (RES) and improve the stability and economy of integrated energy systems (IES). However, the stable demand-supply of IES faces challenges from uncertainties that arise from RES and loads, as well as the increasing impact of cyber-attacks with advanced information and communication technologies adoption. To address these challenges, this paper proposes an innovative model-free resilience scheduling method based on state-adversarial deep reinforcement learning (DRL) for integrated demand response (IDR)-enabled IES. The proposed method designs an IDR program to explore the interaction ability of electricity-gas-heat flexible loads. Additionally, the state-adversarial Markov decision process (SA-MDP) model characterizes the energy scheduling problem of IES under cyber-attack, incorporating cyber-attacks as adversaries directly into the scheduling process. The state-adversarial soft actor-critic (SA-SAC) algorithm is proposed to mitigate the impact of cyber-attacks on the scheduling strategy, integrating adversarial training into the learning process to against cyber-attacks. Simulation results demonstrate that our method is capable of adequately addressing the uncertainties resulting from RES and loads, mitigating the impact of cyber-attacks on the scheduling strategy, and ensuring a stable demand supply for various energy sources. Moreover, the proposed method demonstrates resilience against cyber-attacks. Compared to the original soft actor-critic (SAC) algorithm, it achieves a 10\% improvement in economic performance under cyber-attack scenarios. 
\end{abstract}


\section{Introduction}
\IEEEPARstart{T}{o} address the increasing conflict between energy demand and environmental protection, renewable energy sources (RES) with zero-carbon characteristics have received widespread attention, and countries worldwide are adjusting their energy systems to lessen their reliance on traditional fossil fuels\cite{wang2018review}. Integrated energy systems (IES) can effectively integrate and utilize various energy units, improving the efficiency of energy use and the consumption capacity of intermittent RES, which provides a new solution for improving the ecological environment and achieving a low-carbon sustainable energy system\cite{liu2020region}. However, the coordination of multiple energy units and the multi-uncertainties of RES and loads present challenges for the economic operation of IES\cite{lu2020hydraulic}. In addition, the dependence on advanced information and communication technologies for energy management systems (EMS) has introduced cyber security threats while improving the efficiency, security and reliability of EMS. False data injection attacks (FDIA) are a common type of cyber-attack that can affect and mislead system operators by integrating fake data into real-time instrumentation measurements, leading to erroneous decisions on policy and control\cite{che2018mitigating}. As an example, a cyber-attack in 2015 resulted in a massive power outage in the Ukrainian power grid\cite{liang20162015}. Therefore, it is urgent to study resilient scheduling under cyber-attacks to mitigate the uneconomic operation and reliability problems caused by such attacks.

In the research on energy dispatch in integrated energy systems, relevant measures have been proposed to address the high degree of uncertainty in the system. In \cite{zheng2019stochastic}, the authors used the probability density function to address the uncertainty of RES. Modeling the stochasticity of RES using scenario analysis is also a common method for dealing with uncertainty\cite{emrani2021optimal}. These methods require accurate modeling of renewable energy sources or enumeration scenarios, while demand response (DR) for flexible loads has been shown to be beneficial in addressing the uncertainty of RES. In \cite{li2021optimal}, an integrated demand response (IDR) program is proposed that considers the interaction among electrical, gas, and thermal loads to enhance system economic efficiency. However, current studies often neglect the IDR program under real-time scheduling scenarios. 
Existing literature on FDIA mainly focuses on designing stealthy FDIA, evaluating their impact on systems, and developing strategies to defend against them. A type of cyber-attacks targeting scheduling information in natural gas systems is proposed, and its impact on the operation of the electric and gas energy subsystems  is analyzed\cite{zhao2021coordinated}. Ref. \cite{shayan2019network} proposes a network cyber-secured unit commitment model to improve the economic efficiency of system operation under FDIA. Furthermore, in terms of detecting and filtering bad data, state estimation plays a key role. However, some FDIAs have been deliberately designed to evade detection \cite{liu2016false}. Load redistribution (LR) attack is a specific class of FDIA that involves the use of false load data to impact the system's operation and cause economic losses and physical damages to equipments due to incorrect operational decisions \cite{zhao2021cyber}. Ref. \cite{ding2022cyber} proposes a stealthy heat load redistribution (HLR) attack and analyzes its impact on the security and economics of IES. Given that stealthy FDIAs can bypass state estimation and avoid detection, they can lead to uneconomical system operations. Therefore, it is urgent to study IES resilience scheduling against cyber-attacks.

Deep reinforcement learning (DRL) is increasingly used in real-time energy management for stochastic decision-making without needing prior environmental knowledge\cite{li2023optimal}. It addresses uncertainties in renewable energy sources (RES) and loads using the soft actor-critic (SAC) algorithm\cite{zhang2021soft} and adapts to power fluctuations\cite{li2023deep}. However, these methods don't fully account for cyber-attacks' impact on decision networks. Given the vulnerability of deep neural networks to minor input changes, this study integrates the CROWN-IBP defense training to enhance DRL's robustness against cyber-attacks, supporting the development of resilient scheduling strategies.

\begin{table*}[ht]   
  \centering
  \caption{Comparison of resilience scheduling methods for electric energy systems under cyber-attack scenarios}\label{tab1}
    \begin{tabular}{ccccccc}  
    \specialrule{0.5pt}{0.5pt}{0.5pt}
    \specialrule{0.5pt}{0.5pt}{0.5pt}
    Reference & Cyber-attack & \multicolumn{2}{c}{Muti-uncertainties} & DR & Test system & Solving method \\
                                &       & RES & Load    &     &      &         \\
    \midrule
    \cite{tajalli2020resilient} & DOSA  & $\times $ & $\times $ & $\times $ & Microgrid & Average consensus-based algorithm  \\  \cite{chu2022mitigating}    & LAA   & $\times $ & $\times $ & $\times $ & Power grid & \begin{tabular}[c]{@{}c@{}}LAA detection + IBR droop control+ \\ Distributionally robust method\end{tabular}   \\
    \cite{huang2021distributed} & Noncolluding and colluding attack   & $\times $ & $\times $ & $\times $ & IES (electricity-heat) &  \begin{tabular}[c]{@{}c@{}}Attack detection and isolation+ \\ Distributionally robust method\end{tabular}   \\
    \cite{zhao2021cyber}        & FDIA (LR) & $\surd $ & $\times $ & $\times $ & MEDS & DRO  \\
    \cite{zhao2020cyber}        & FDIA(LR) & $\surd $ & $\times $ & $\times $ & WES  & DRO   \\
    \cite{hua2019optimal}        & $\times $ & $\surd $ & $\surd $ & $\times $ & Microgrid  & DRL algorithm   \\
    Proposed                    & FDIA(HLR) & $\surd $ & $\surd $ & $\surd $ & IES (electricity-gas-heat)      & DRL algorithm   \\ 
    \specialrule{0.5pt}{0.5pt}{0.5pt}
    \specialrule{0.5pt}{0.5pt}{0.5pt}
    \end{tabular}%
  \label{tab:addlabel}%
\end{table*}%

Table \ref{tab1} compares our proposed approach  with current resilient scheduling methods for cyber-attack scenarios, emphasizing its distinctive features. In the table, 
“DRO” refers to the distributionally robust optimization; “DOSA” refers to the denial of service attack; “LAA” denotes the load-altering attack; “MEDS” represents the multivector energy distribution system; “WES” denotes the water-energy system; “IBR” refers to the inverter-based resource. 

The existing research gaps in electric energy system scheduling considering cyber-attack can be summarized as: (1) Dealing with multi-uncertainties of RES and loads simultaneously is challenging, (2) DR problem and the potential of flexible resources for resilient IES has not been fully exploited, and (3) To the authors' knowledge, few studies use DRL-based methods for IES resilient scheduling against cyber-attacks.

The main contributions of this work are summarized below:

(1) A resilient scheduling method for IES is developed to counter cyber-attacks, addressing IES's vulnerability in such scenarios. It analyzes the role of energy storage and DR in mitigating attacks and introduces a dynamic pricing mechanism that combines time-of-use (TOU) and real-time pricing, considering user thermal comfort through the predicted average voting index. This method, distinct from traditional ones, employs a DRL agent to establish optimal pricing strategies under source-load uncertainties. 

(2) We consider the uncertainty of state observation in reinforcement learning, take the cyber-attack suffered by IES as the uncertain input of state observation, and transform the dynamic energy scheduling problem of IES under cyber-attack into a state adversarial Markov decision process (SA-MDP). The SA-MDP model handles continuous state and action spaces to reflect the nature of the IES problem.

(3) A proposed DRL algorithm called state-adversarial soft actor-critic (SA-SAC) adds a regularizer to the policy network during training to reflect the hazard degree of cyber-attacks. A certified defense training method is employed to train policy neural network, aiming to identify the optimal learning policy that demonstrates resilience against cyber-attacks. The superiority of the proposed algorithm is verified in the scenario where cyber-attacks are used as state observation adversary. 

\section{Modeling of IES and LR Attacks}
\subsection{Structure Modeling of IES}
\begin{figure}[htpb]
    \centering
    \includegraphics[width=3.2in,height=2.2in]{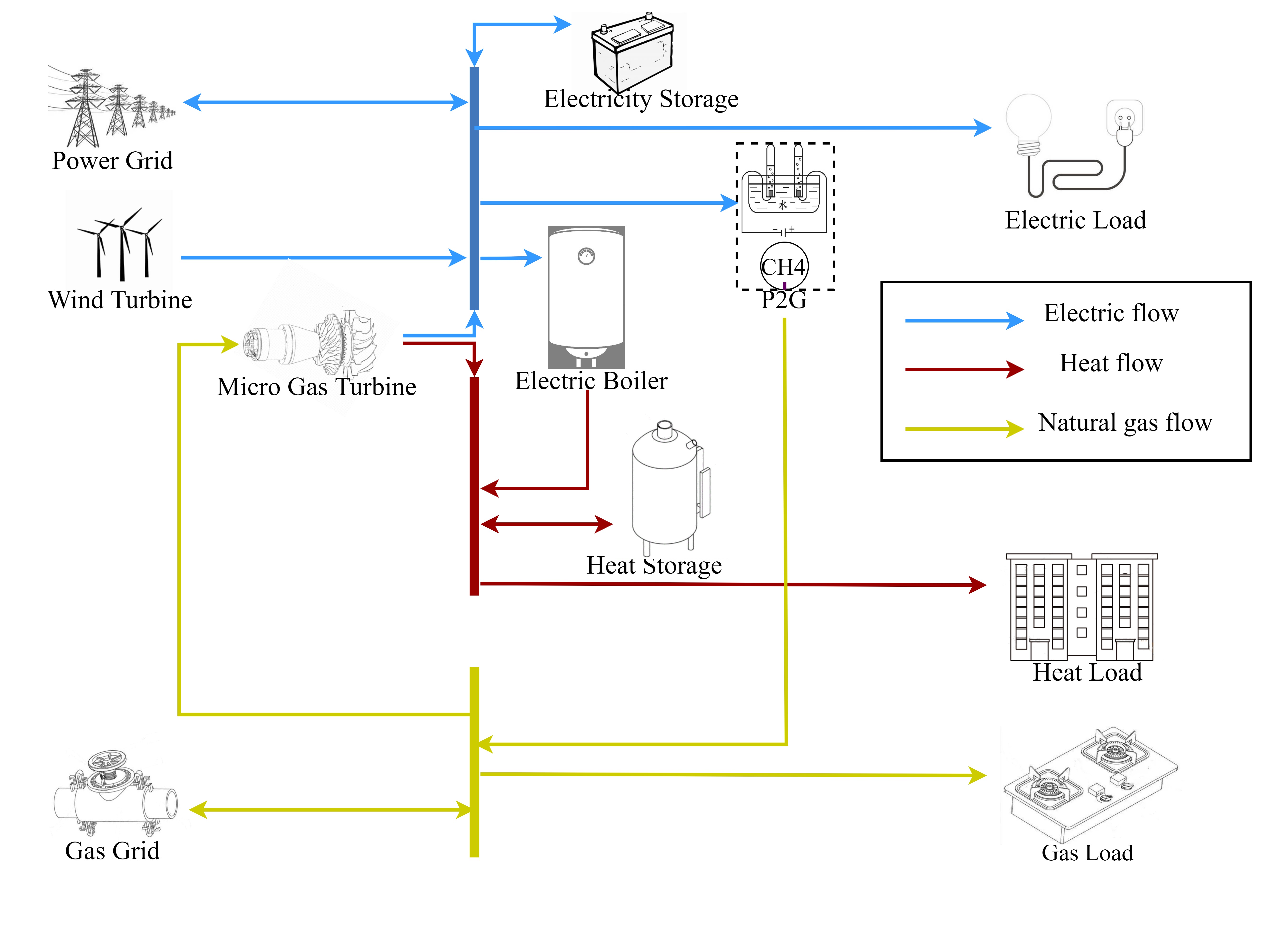}
    \caption{Schematic diagram of the IES.}
    \label{CIES}
\end{figure}

The overall structure of an IES is shown in Fig. \ref{CIES}, including micro-gas turbine (MT), wind turbine (WT), electricity storage device (ESD), heat storage device (HSD), electric boiler (EB), power to gas (P2G), electric load, heat load and gas load. The system integrates and distributes electricity, heat, and natural gas to customers via electric power, district heating, and natural gas networks. 
\subsubsection{Modeling of Load}
\paragraph{Electric Load Model}
The study classifies electrical loads into fixed and flexible loads. Flexible loads, primarily time-shifted electrical loads, have fixed power consumption but flexible power consumption time. Users receive subsidies to compensate for the impact of load shifting. The actual electrical load, time-shifted electrical load, and reimbursed electrical load at period $t$ are 
{\setlength\abovedisplayskip{2pt}
\setlength\belowdisplayskip{2pt}
\begin{equation}\label{Eq1}
{{P}_{load,t}}={{P}_{basic,t}}-{{P}_{TSE,t}}+{{P}_{PBE,t}}\quad \forall t
\end{equation} }{\setlength\abovedisplayskip{2pt}
\setlength\belowdisplayskip{2pt}
\begin{equation}\label{Eq2}
{{P}_{TSE,t}}={{P}_{basic,t}}*\sigma _{t}^{TSE}*\varsigma _{t}^{TSE}\quad \forall t
\end{equation} }{\setlength\abovedisplayskip{2pt}
\setlength\belowdisplayskip{2pt}
\begin{equation}\label{Eq3}
{{P}_{PBE,t}}=\sum\limits_{i=1}^{t-1}({\omega _{i,t}^{PBE}}*{{P}_{TSE,t}})\quad \forall t
\end{equation} }where, in period $t$, ${{P}_{basic,t}}$ and ${{P}_{load,t}}$ represent initial and actual electrical loads; ${{P}_{TSE,t}}$ and ${{P}_{PBE,t}}$ denote the actual time-shifted load amount and actual load repayment amount. $\varsigma_{t}^{TSE}$ indicates price level, $\sigma_{t}^{TSE}$ reflects load change sensitivity to price. The binary variable $\omega_{i,t}^{PBE}$, determined by the load transfer probability ${{Pr}_{PBE}}$, signifies the repayment of the time-shifted electrical load, with 1 indicating repayment and 0 indicating no repayment\cite{nakabi2021deep}.
{\setlength\abovedisplayskip{2pt}
\setlength\belowdisplayskip{2pt}
\begin{equation}\label{Eq4}
{{Pr}_{PBE,t}}=clip\left( \frac{-\varsigma _{t}^{TSE}*sign\left( {{P}_{TSE,t}} \right)}{2}+\frac{t-i}{{{\eta }_{TSE}}},0,1 \right)\text{ }\forall t\text{ }
\end{equation} }{\setlength\abovedisplayskip{2pt}
\setlength\belowdisplayskip{2pt}
\begin{equation}\label{Eq5}
clip(X,0,1)=\left\{ \begin{array}{*{35}{l}}
   \text{ }0, & \text{ if   }X<0  \\
   X, & \text{ if   }0\le X\le 1  \\
   \text{ }1, & \text{ if   }X>1  \\
\end{array} \right.
\end{equation} }where the probability $Pr_{PBE}$ relies on the price level ${{\varsigma }_{t}}$ and the patience factor $\eta $; $clip()$ limits the load transfer probability. The compensation of load transfer is governed by probabilistic functions associated with the "patience factor," "price levels," and the temporal dynamics within the scheduling cycle. This factor impacts the probability and timing of load compensation, ensuring that any load shifted from its initial timing is fully compensated within subsequent operational cycles.

\paragraph{Gas Load Model}
The actual gas load, time-shifted gas load, and reimbursed gas load at period $t$ are formulated as follows:
{\setlength\abovedisplayskip{2pt}
\setlength\belowdisplayskip{2pt}
\begin{equation}\label{Eq6}
{{Q}_{load,t}}={{Q}_{basic,t}}-{{Q}_{TSQ,t}}+{{Q}_{PBQ,t}}\quad \forall t
\end{equation} }{\setlength\abovedisplayskip{2pt}
\setlength\belowdisplayskip{2pt}
\begin{equation}\label{Eq7}
{{Q}_{TSQ,t}}={{Q}_{basic,t}}*\sigma _{t}^{TSQ}*\varsigma _{t}^{TSQ}\quad \forall t
\end{equation} }{\setlength\abovedisplayskip{2pt}
\setlength\belowdisplayskip{2pt}
\begin{equation}\label{Eq8}
{{Q}_{PBQ,t}}=\sum\limits_{i=1}^{t-1}({\omega _{i,t}^{PBQ}}*{{Q}_{TSQ,t}})\quad \forall t
\end{equation} }where ${{Q}_{basic,t}}$ and ${{Q}_{load,t}}$ are the initial and actual gas loads in period $t$; ${{Q}_{TSQ,t}}$ is the time-shifted gas load value; ${{Q}_{PBQ,t}}$ is the time-shifted gas load payback value; $\varsigma _{t}^{TSQ}$ is the price level that influences the magnitude of the time-shifted gas load; $\sigma _{t}^{TSQ}$ is the sensitivity factor, which represents the percentage of gas load change as the price fluctuates; $\omega _{i,t}^{PBQ}$ is a binary variable that represents the repayment of the time-shifted gas load and is determined by the load transfer probability ${{Pr}_{PBQ}}$.

{\setlength\abovedisplayskip{2pt}
\setlength\belowdisplayskip{2pt}
\begin{equation}\label{Eq9}
{{Pr}_{PBQ,t}}=\operatorname{clip}\left( \frac{-\varsigma _{t}^{TSQ}*\operatorname{sign}\left( {{Q}_{TSQ,t}} \right)}{2}+\frac{t-i}{{{\eta }_{TSQ}}},0,1 \right)\quad \forall t\text{ }
\end{equation} }
\paragraph{Heat Load Model}
Considering the thermal inertia of the building in the heat load model, the heating power stored in buildings can be calculated via the first-order thermodynamic model. In addition, users have some flexibility in their temperature perception, which can be utilized to reduce the thermal load while still maintaining their preferred thermal comfort level. Based on this, the actual thermal load and the reduced thermal load are  
{\setlength\abovedisplayskip{2pt}
\setlength\belowdisplayskip{2pt}
\begin{equation}\label{Eq12}
{{H}_{load,t}}={{H}_{basic,t}}-{{H}_{CH,t}}\quad \forall t
\end{equation} }{\setlength\abovedisplayskip{2pt}
\setlength\belowdisplayskip{2pt}
\begin{equation}\label{Eqadd3}
H_{C H, t}=\varsigma_t^{CH} H_{C H, t}^{\max } \quad \forall t
\end{equation} }{\setlength\abovedisplayskip{2pt}
\setlength\belowdisplayskip{2pt}
\begin{equation}\label{Eq13}
0\le {{H}_{CH,t}}\le H_{CH,t}^{\max }
\end{equation} }where ${{H}_{load,t}}$ denotes the actual thermal load in period $t$; ${{H}_{basic,t}}$ denotes the initial thermal load in period $t$; ${{H}_{CH,t}}$ denotes the users’ cuttable thermal load in period $t$; $H_{CH,t}^{\max }$is the maximum cuttable thermal power; $\varsigma _{t}^{CH}$ denotes the price level that influences the amount of the cuttable thermal load.

To quantify the acceptable user thermal comfort, the predicted mean vote (PMV) index is introduced as follows\cite{mao2019pmv}:
{\setlength\abovedisplayskip{2pt}
\setlength\belowdisplayskip{2pt}
\begin{equation}\label{Eq14}
P M V=2.43-\frac{3.76\left(T_s-T_{i n, t}\right)}{M\left(I_{c l}+0.1\right)}
\end{equation} }where ${{I}_{cl}}$ stands for the thermal resistance of clothing; ${{T}_{in,t}}$ denotes the indoor temperature. ${{T}_{s}}$ is the average temperature at which a person's skin feels comfortable; $M$ denotes the human body's metabolic rate. Here, the indoor comfort temperature is taken as 21 degrees Celsius. 

As a vital indicator of comfort, the PMV index encapsulates users' subjective perceptions of indoor temperatures, which in turn informs scheduling decisions. To ensure indoor thermal comfort, it is recommended that the PMV value stays within the range of [-0.5, 0.5] \cite{ku2014automatic}. In this study, we limit the nighttime PMV to the range of [-0.9, 0.9] \cite{li2021coordinating}. The PMV change range is 

{\setlength\abovedisplayskip{2pt}
\setlength\belowdisplayskip{2pt}
\begin{equation}\label{Eq15}
\begin{cases}|P M V| \leq 0.9, & {[1: 00-7: 00] \cup [20: 00-24: 00]} \\ |P M V| \leq 0.5, & {[8: 00-19: 00]}\end{cases}
\end{equation} }
Additionally, the thermal load is predominantly utilized to sustain indoor temperatures by transferring heat via a network of heat pipes, thereby fulfilling the heating demands of the building. A constant mass flow rate is maintained to control the water temperature within the heating system, effectively satisfying the thermal load requirements. The transient heat balance equation of the building is employed to depict the impact of heat variations in the heating system on the building temperature \cite{li2021optimal}. This equation allows us to calculate the building's heat load:
{\setlength\abovedisplayskip{2pt}
\setlength\belowdisplayskip{2pt}
\begin{equation}\label{Eq11}
{{H}_{basic,t}}=\frac{\left[ {{T}_{in,t}}-{{T}_{\text{out },t}} \right]+\frac{K\cdot F}{{{c}_{\text{air }}}\cdot {{\rho }_{\text{air }}}\cdot V}\cdot \Delta t\cdot \left[ {{T}_{in,t-1}}-{{T}_{\text{out },t}} \right]}{\frac{1}{K\cdot F}+\frac{1}{{{c}_{\text{air }}}\cdot {{\rho }_{\text{air }}}\cdot V}\cdot \Delta t}
\end{equation} }where ${{T}_{out,t}}$ represents the outdoor temperature in period $t$; $K$ denotes the comprehensive heat transfer coefficient; $F$ and $V$ denote the surface area and volume of the building, respectively; ${{c}_{air}}$ and ${{\rho }_{air}}$ refer to the specific heat capacity and density of the indoor air, respectively.

\subsubsection{Energy Storage Device Model}
This study categorizes the energy storage devices discussed into two types: ESD and HSD. The ESD and HSD models are formulated as:
{\setlength\abovedisplayskip{2pt}
\setlength\belowdisplayskip{2pt}
\begin{equation}\label{Eq16}
\left\{\begin{array}{l}
C_{t+1}^{E S D}=C_t^{E S D}+\left(\eta_{c h}^{E S D} P_{c h, t}^{E S D}-P_{d c, t}^{E S D} / \eta_{d c}^{E S D}\right) \Delta t \\
C_{t+1}^{H S D}=C_t^{H S D}+\left(\eta_{c h}^{H S D} H_{c h, t}^{H S D}-H_{d c, t}^{H S D} / \eta_{d c}^{H S D}\right) \Delta t
\end{array} \forall t\right.
\end{equation} }where $P_{ch,t}^{ESD}$ and $P_{dc,t}^{ESD}$ denote the charging and discharging powers of the ESD in period $t$, respectively; $H_{ch,t}^{HSD}$ and $H_{dc,t}^{HSD}$ denote the HSD heat storage and release powers; $\eta _{ch}^{ESD}$ and $\eta _{dc}^{ESD}$ are the ESD charge and discharge efficiencies, respectively; $\eta _{ch}^{HSD}$ and $\eta _{dc}^{HSD}$ are the HSD heat storage and release efficiencies; $C_{t}^{ESD}$ and $C_{t}^{HSD}$ are the capacity values of the ESD and HSD. 
The state of charge (SOC) of the energy storage device is  
{\setlength\abovedisplayskip{2pt}
\setlength\belowdisplayskip{2pt}
\begin{equation}\label{Eq17}
\left\{\begin{array}{l}
S O C_t^{E S D}=C_t^{E S D} / C_{\max }^{E S D} \\
S O C_t^{H S D}=C_t^{H S D} / C_{\max }^{H S D}
\end{array} \forall t\right.
\end{equation} }where $C_{\max }^{ESD}$ and $C_{\max }^{HSD}$ denote the maximum capacity of the ESD and HSD, respectively.

\subsubsection{Electric Boiler Model}

The operation of the EB is modelled as
{\setlength\abovedisplayskip{2pt}
\setlength\belowdisplayskip{2pt}
\begin{equation}\label{Eq18}
{{H}_{EB,t}}={{\eta }^{EB}}{{P}_{EB,t}}\quad \forall t
\end{equation} }where ${{P}_{EB,t}}$ denotes the absorbed electrical energy in period $t$; ${{H}_{EB,t}}$ denotes the heat output of the EB in period $t$; ${{\eta }^{EB}}$ is the conversion efficiency of the EB.

\subsubsection{Power to Gas Model}
Our model employs P2G technology to convert electrical energy into gas, enhancing flexibility and managing the intermittency of renewable sources like wind and solar power. The natural gas production by P2G conversion is correlated with the electricity consumption in the following manner: 
{\setlength\abovedisplayskip{2pt}
\setlength\belowdisplayskip{2pt}
\begin{equation}\label{Eq19}
{{Q}_{P2G,t}}=\frac{{{\eta }_{P2G}}{{P}_{P2G,t}}}{HHV}\quad \forall t
\end{equation} }where ${{Q}_{P2G,t}}$ denotes the natural gas produced by P2G in period $t$; ${{P}_{P2G,t}}$ denotes the absorbed electrical energy in period $t$; $HHV$ is the calorific value of natural gas; ${{\eta }_{P2G}}$ is the conversion efficiency of P2G.

\subsubsection{Micro-Gas Turbine Model}

The output powers of the MT are formulated by
{\setlength\abovedisplayskip{2pt}
\setlength\belowdisplayskip{2pt}
\begin{equation}\label{Eq20}
{{P}_{MT,t}}=(1-\eta _{h}^{MT}-\eta _{loss}^{MT}){{Q}_{MT,t}}HHV\quad \forall t
\end{equation} }{\setlength\abovedisplayskip{2pt}
\setlength\belowdisplayskip{2pt}
\begin{equation}\label{Eq21}
{{H}_{MT,t}}=\left( 1-\eta _{e}^{MT}-\eta _{\text{loss }}^{MT} \right){{Q}_{MT,t}}HHV\quad \forall t
\end{equation} }where ${{H}_{MT,t}}$ is the heat output of the MT in period $t$; ${{P}_{MT,t}}$ is the electrical power output of the MT; ${{Q}_{MT,t}}$ is the volume of natural gas consumed by the MT; $\eta _{e}^{MT}$ and $\eta _{h}^{MT}$ are the electric power and heat generation efficiency coefficients of the MT; $\eta _{loss}^{MT}$ is the heat loss coefficient. 
\vspace{-3mm}
\subsection{Modeling of LR Attack}

LR attack is a type of FDIA that alters load data, leading to incorrect meter readings and misleading system operators in making scheduling decisions \cite{zhao2021cyber}.

\subsubsection{Bad Data Detection in IES State Estimation}

The state estimation in an IES is expressed as follows\cite{zang2019robust}:
{\setlength\abovedisplayskip{2pt}
\setlength\belowdisplayskip{2pt}
\begin{equation}\label{Eq22}
z=h(x)+e
\end{equation} }where $z\in \left\{ {{z}_{e}},{{z}_{g}},{{z}_{h}} \right\}$ is the measurement variables in the electricity, gas and heat networks; $x\in \left\{ {{x}_{e}},{{x}_{g}},{{x}_{h}} \right\}$ represents the state variables; $h(x)$ denotes the measurement function of $x$; $e$ is the measurement noise.

The common bad data detection method is based on the residual test principle, aiming to minimize the measurement residual $r$, which is defined as:
{\setlength\abovedisplayskip{2pt}
\setlength\belowdisplayskip{2pt}
\begin{equation}\label{Eq23}
||r|{{|}_{2}}=||z-h(x)|{{|}_{2}}
\end{equation} }

One can utilize the largest normalized residual to detect and recognize measurement errors, and we evaluate the Euclidean norm of the residual against a predefined threshold value $\tau$:
{\setlength\abovedisplayskip{2pt}
\setlength\belowdisplayskip{2pt}
\begin{equation}\label{Eq24}
||r|{{|}_{2}}=||z-h(x)|{{|}_{2}}\le \tau
\end{equation} }

If the residual is less than the threshold, the state estimate is valid; otherwise, it indicates the presence of bad data. 

\subsubsection{HLR Attack Modeling}
The less mature state estimation of the heating system makes it a vulnerable spot in the IES. The "barrel principle" means that a single subsystem failure may propagate to other energy subsystems. Due to the interdependency of the thermal system with the electrical and natural gas systems, any attacks on the thermal load could indirectly influence the functionality of these systems via interconnected devices, potentially precipitating a more extensive chain reaction. Thus, the analysis of HLR attacks and their impacts on various subsystems is essential for understanding these intricate chain reactions. This study focuses on a stealthy HLR attack, the indoor temperature delay scaling attack (ITDSA), notable for its economic impact. The HLR attack is a sophisticated cyber-attack strategy, wherein attackers exploit system vulnerabilities to skillfully manipulate indoor temperature data, thereby influencing system operations. This type of attack has gained recognition in the literature for its potential to disrupt integrated energy systems by altering critical data, and is notably challenging to detect\cite{ding2022cyber}. The ITDSA is characterized as follows:

{\setlength\abovedisplayskip{2pt}
\setlength\belowdisplayskip{2pt}
\begin{equation}\label{Eq25}
\tilde{\tau }_{in}^{t}=\left\{ \begin{array}{*{35}{l}}
   \tau _{in}^{t} & t\notin {{\Gamma }^{a}}  \\
   \left( 1+\lambda \left( 1-{{e}^{-a\left( t-{{t}_{0}} \right)}} \right) \right)\tau _{in}^{t} & t\in {{\Gamma }^{a}}  \\
\end{array} \right.
\end{equation} }where $\tau _{in}^{t}$ is the indoor temperature; $\lambda $ and $a$ are the parameters of the HLR attack. According to Eq. (\ref{Eq11}), directly manipulating the indoor temperature results in a corresponding alteration in the heat load. 

\vspace{-2mm}
\section{Formulation of IES scheduling model}

In this section, we formulate the IES dispatch model that pursues the maximum net profit for the operator. A dynamic pricing mechanism that integrates TOU and real-time prices is proposed to fully exploit the demand response potential.
\vspace{-3mm}
\subsection{Dynamic Pricing Mechanism}

Our dynamic pricing method integrates DRL to manage electricity, gas, and heat prices. Using TOU prices as a benchmark, it offers customers predictable pricing, while allowing the supplier to adjust prices in real-time based on grid feedback and IES device status. This mechanism guides customer demand and increases IES operator profits. Dynamic price limits, tied to TOU price peaks and valleys, ensure balanced prices without extremes. The mechanism is 
{\setlength\abovedisplayskip{2pt}
\setlength\belowdisplayskip{2pt}
\begin{equation}\label{Eq26}
\left\{\begin{array}{l}
\lambda_{rt, t}^P=\varpi_{P, m}+\varsigma_t^{TSE} * k_P \\
\lambda_{rt, t}^Q=\varpi_{Q, m}+\varsigma_t^{TSQ} * k_Q \quad \forall t \\
\lambda_{rt, t}^H=\varpi_{H, m}+\varsigma_t^{CH} * k_H
\end{array}\right.
\end{equation} }{\setlength\abovedisplayskip{2pt}
\setlength\belowdisplayskip{2pt}
\begin{equation}\label{Eq27}
\left\{\begin{array}{l}
\varpi_{P, m}=\zeta_P\left(\varpi_{s t, u}^P+\varpi_{s t, d}^P\right) \\
\varpi_{Q, m}=\zeta_Q\left(\varpi_{s t, u}^Q+\varpi_{s t, d}^Q\right) \quad \forall t \\
\varpi_{H, m}=\varpi_{s t, t}^H
\end{array}\right.
\end{equation} }{\setlength\abovedisplayskip{2pt}
\setlength\belowdisplayskip{2pt}
\begin{equation}\label{Eq28}
\left\{\begin{array}{c}
l_P \varpi_{s t, d}^P \leq \lambda_{r t, t}^P \leq l_P \varpi_{s t, u}^P \\
l_Q \varpi_{s t, d}^Q \leq \lambda_{r t, t}^Q \leq l_Q \varpi_{s t, u}^Q \\
\varpi_{s t, t}^H \leq \lambda_{r t, t}^H \leq l_H \varpi_{s t, t}^H
\end{array} \forall t\right.
\end{equation} }where $\varpi _{st,t}^{P}$, $\varpi _{st,t}^{Q}$ and $\varpi _{st,t}^{H}$ are the TOU prices of electricity, gas and heat subsystems, respectively; $\varpi_{s t, u}^P$, $\varpi_{s t, u}^Q$ and $\varpi_{s t, d}^P$, $\varpi_{s t, d}^Q$ are the peak and off-peak values of TOU prices of electricity and gas subsystems; $\lambda _{rt,t}^{P}$, $\lambda _{rt,t}^{Q}$ and $\lambda _{rt,t}^{H}$ are the real-time electricity, gas and heat prices issued by the operator to users; ${{\varpi }_{P,m }}$, ${{\varpi }_{Q,m }}$, and ${{\varpi }_{H,m }}$ are the benchmark prices for electricity, gas, and heat subsystem; ${{l}_{P}}$, ${{l}_{Q}}$, and ${{l}_{H}}$ denote the threshold coefficients for real-time electricity, gas and heat prices; ${{k}_{P}}$, ${{k}_{Q}}$ and ${{k}_{H}}$ are the parameters that determine the change in the prices of electricity, gas and heat.

\vspace{-2mm}
\subsection{Objective Function}
The objective function aims to maximize the profit of the operator and is given by:
{\setlength\abovedisplayskip{2pt}
\setlength\belowdisplayskip{2pt}
\begin{equation}\label{Eq29}
\max F=\sum_{t=1}^T\left(Re{{v}_{t}}-Cos{{t}_{t}}\right)
\end{equation} }{\setlength\abovedisplayskip{2pt}
\setlength\belowdisplayskip{2pt}
\begin{equation}\label{Eq30}
\begin{aligned}
Re{{v}_{t}} & =\lambda_{rt, t}^P \sum_{\text {loads }} P_{\text {load }, t}+\lambda_{rt, t}^Q \sum_{\text {loads }} Q_{\text {load }, t}+\lambda_{rt, t}^H \sum_{\text {loads }} H_{\text {load }, t} \\
& +\lambda_{\text {sell }, t}^P P_{\text {sell }, t}+\lambda_{\text {sell }, t}^Q Q_{\text {sell }, t}
\end{aligned}
\end{equation} }{\setlength\abovedisplayskip{2pt}
\setlength\belowdisplayskip{2pt}
\begin{equation}\label{Eq31}
\begin{aligned}
Co{{s}_{t}} & =P_{b u y, t} \lambda_{\text {buy }, t}^P+Q_{b u y, t} \lambda_{b u y, t}^Q+\gamma_{C H} H_t^{C H} \\
& +\gamma_{T S E} \sum_{T S E} \max \left(P_t^{T S E}, 0\right) \\
& +\gamma_{T S Q} \sum_{T S Q} \max \left(P_t^{T S Q}, 0\right)
\end{aligned}
\end{equation} }where $Re{{v}_{t}}$ and $Co{{s}_{t}}$ are the total income and the operating cost of the operator; $\lambda _{buy,t}^{P}$ and $\lambda _{sell,t}^{P}$ are the unit prices of electricity purchased and sold by the IES from and to the power grid; $\lambda _{buy,t}^{Q}$ and $\lambda _{sell,t}^{Q}$ are the unit prices of natural gas purchased and sold by the IES from and to the gas grid; ${{P}_{buy,t}}$ and ${{P}_{sell,t}}$ are the electricity purchased and sold by the IES from and to the power grid, respectively; ${{Q}_{buy,t}}$ and ${{Q}_{sell,t}}$ are the natural gas purchased and sold by the IES from and to the gas grid; ${{\gamma }_{TSE}}$ and ${{\gamma }_{TSQ}}$ are the unit compensation cost of time-shifted electric load and time-shifted gas load; ${{\gamma }_{CH}}$ is the unit compensation cost of cutting heat load.
\vspace{-4mm}
\subsection{Constraint Conditions}
\subsubsection{Energy Balance Constraints}
The energy balance constraints are given below: 
{\setlength\abovedisplayskip{2pt}
\setlength\belowdisplayskip{2pt}
\begin{equation}\label{Eq32}
{{P}_{\text{grid,t }}}+{{P}_{WT,t}}+{{P}_{ESD,t}}+{{P}_{MT,t}}-{{P}_{EB,t}}-{{P}_{P2G,t}}={{P}_{\text{load,t }}}\text{  }\forall t
\end{equation} }{\setlength\abovedisplayskip{2pt}
\setlength\belowdisplayskip{2pt}
\begin{equation}\label{Eq33}
{{H}_{MT,t}}+{{H}_{HSD,t}}+{{H}_{EB,t}}={{H}_{\text{load,t }}}\text{  }\forall t
\end{equation} }{\setlength\abovedisplayskip{2pt}
\setlength\belowdisplayskip{2pt}
\begin{equation}\label{Eq34}
{{Q}_{\text{grid,t }}}+{{Q}_{P2G,t}}-{{Q}_{MT,t}}={{Q}_{\text{load,t }}}\text{  }\forall t
\end{equation} }where ${{P}_{WT,t}}$ is the output of the WT; ${{P}_{ESD,t}}$ and ${{H}_{HSD,t}}$ are the charging/discharging power of ESD and HSD, respectively.

\subsubsection{ESD Operating Constraints}
In order to ensure that the charging and discharging powers of the ESD are within the allowable ranges, the charging and discharging power are constrained as:
{\setlength\abovedisplayskip{2pt}
\setlength\belowdisplayskip{2pt}
\begin{equation}\label{Eq35}
\left\{\begin{array}{l}
0 \leq P_{d c, t}^{E S D} \leq \theta_{d c, t}^{E S D} P_{d c, \text { max }}^{E S D} \\
0 \leq P_{c h, t}^{E S D} \leq \theta_{c h, t}^{E S D} P_{c h, \max }^{E S D}
\end{array} \forall t\right.
\end{equation} }{\setlength\abovedisplayskip{2pt}
\setlength\belowdisplayskip{2pt}
\begin{equation}\label{Eq36}
\theta _{dc,t}^{ESD}+\theta _{ch,t}^{ESD}\le 1
\end{equation} }where $P_{ch,\max }^{ESD}$ and $P_{dc,\max }^{ESD}$ are the maximum charging and discharging powers of the ESD, respectively; $\theta _{ch,t}^{ESD}$ and $\theta _{dc,t}^{ESD}$ are Boolean variables that represent the ESD running states.

To ensure consistent initial conditions and within-range capacity in each cycle, the ESD capacity must satisfy:
{\setlength\abovedisplayskip{2pt}
\setlength\belowdisplayskip{2pt}
\begin{equation}\label{Eq37}
\left\{\begin{array}{l}
C_{\min }^{E S D} \leq C_t^{E S D} \leq C_{\max }^{E S D} \\
C_0^{E S D}=C_{\text {end }}^{E S D}
\end{array} \forall t\right.
\end{equation} }where $C_{\min }^{ESD}$ and $C_{\max }^{ESD}$ refer to the ESD's minimum and maximum capacities; $C_{0}^{ESD}$ and $C_{end}^{ESD}$ are the starting and ending capacities of the ESD, respectively.There are similar operational constraints for HSD which are not restated here.

\subsubsection{Electric Boiler Constraint}
The output power of the EB must meet the constraint:
{\setlength\abovedisplayskip{2pt}
\setlength\belowdisplayskip{2pt}
\begin{equation}\label{Eq41}
0\le {{H}_{EB,t}}\le {{H}_{EB,\max }}\text{ }\forall t
\end{equation} }where ${{H}_{EB,\max }}$ is the maximum heat power output of EB.

\subsubsection{P2G Operating Constraints}
The inputs to P2G obey the following constraints:
{\setlength\abovedisplayskip{2pt}
\setlength\belowdisplayskip{2pt}
\begin{equation}\label{Eq42}
P_{P 2 G, \text { min }} \leq P_{P 2 G, t} \leq P_{P 2 G, \text { max }} \forall t
\end{equation} }{\setlength\abovedisplayskip{2pt}
\setlength\belowdisplayskip{2pt}
\begin{equation}\label{Eq43}
\Delta P_{P 2 G, \min } \leq P_{P 2 G, t}-P_{P 2 G, t-1} \leq \Delta P_{P 2 G, \max } \forall t
\end{equation} }where ${{P}_{P2G,\max }}$ and ${{P}_{P2G,\min }}$ are the maximum and minimum input power of P2G, respectively; $\Delta P_{P 2 G, \max }$ and $\Delta {{P}_{P2G,\min }}$ are the upper and lower limits of the climbing ability of P2G, respectively.

\subsubsection{MT Operating Constraints}
The inputs to MT obey the following constraints:
{\setlength\abovedisplayskip{2pt}
\setlength\belowdisplayskip{2pt}
\begin{equation}\label{Eq44}
{{Q}_{MT,\min }}\le {{Q}_{MT,t}}\le {{Q}_{MT,\max }}\text{  }\forall t
\end{equation} }{\setlength\abovedisplayskip{2pt}
\setlength\belowdisplayskip{2pt}
\begin{equation}\label{Eq45}
\Delta {{Q}_{MT,\min }}\le {{Q}_{MT,t}}-{{Q}_{MT,t-1}}\le \Delta {{Q}_{MT,\max }}\text{  }\forall t
\end{equation} }where ${{Q}_{MT,\max }}$ and ${{Q}_{MT,\min }}$ are the maximum and minimum gas consumption volume of MT, respectively; $\Delta {{Q}_{MT,\max }}$ and $\Delta {{Q}_{MT,\min }}$ are the upper and lower limits of the climbing ability of MT, respectively.

\subsubsection{External Energy Transmission Power Constraints}
The energy trading limit between the integrated energy system and the power grid and gas grid can be expressed as:
{\setlength\abovedisplayskip{2pt}
\setlength\belowdisplayskip{2pt}
\begin{equation}\label{add1}
\left\{\begin{array}{l}
P_{\text { min }}^{\text {grid }} \leq P_{t}^{\text {grid }} \leq P_{ \text { max }}^{\text {grid }} \\
Q_{\text { min }}^{\text {grid }} \leq Q_{t}^{\text {grid }} \leq Q_{\text { max }}^{\text {grid }}
\end{array}\right.
\end{equation} }where $P_{\text{max }}^{\text{grid }}$ and $P_{\text{min }}^{\text{grid }}$ are the upper and lower limits of the interactive electric power between IES and the power grid, respectively; $Q_{\text{ max}}^{\text{grid }}$ and $Q_{\text{min }}^{\text{grid }}$ are the upper and lower limits of IES and gas grid interactive natural gas, respectively.

\section{Model solving}
This section presents a SA-MDP model and a DRL algorithm for resilient scheduling decisions. The SA-MDP model aims to maximize operator profits while balancing supply and demand, considering load and RES uncertainties, and the impact of HLR attack.

\subsection{SA-MDP Model}
To maximize intra-day operator profits and consider cyber-attacks, the IES scheduling problem is formulated as a SA-MDP. It incorporates constraints and uncertainties from different sources like MT, P2G, WT, and fluctuating loads. 

\begin{figure}[ht]
    \centering
    \includegraphics[width=3.6in,height=1.6in]{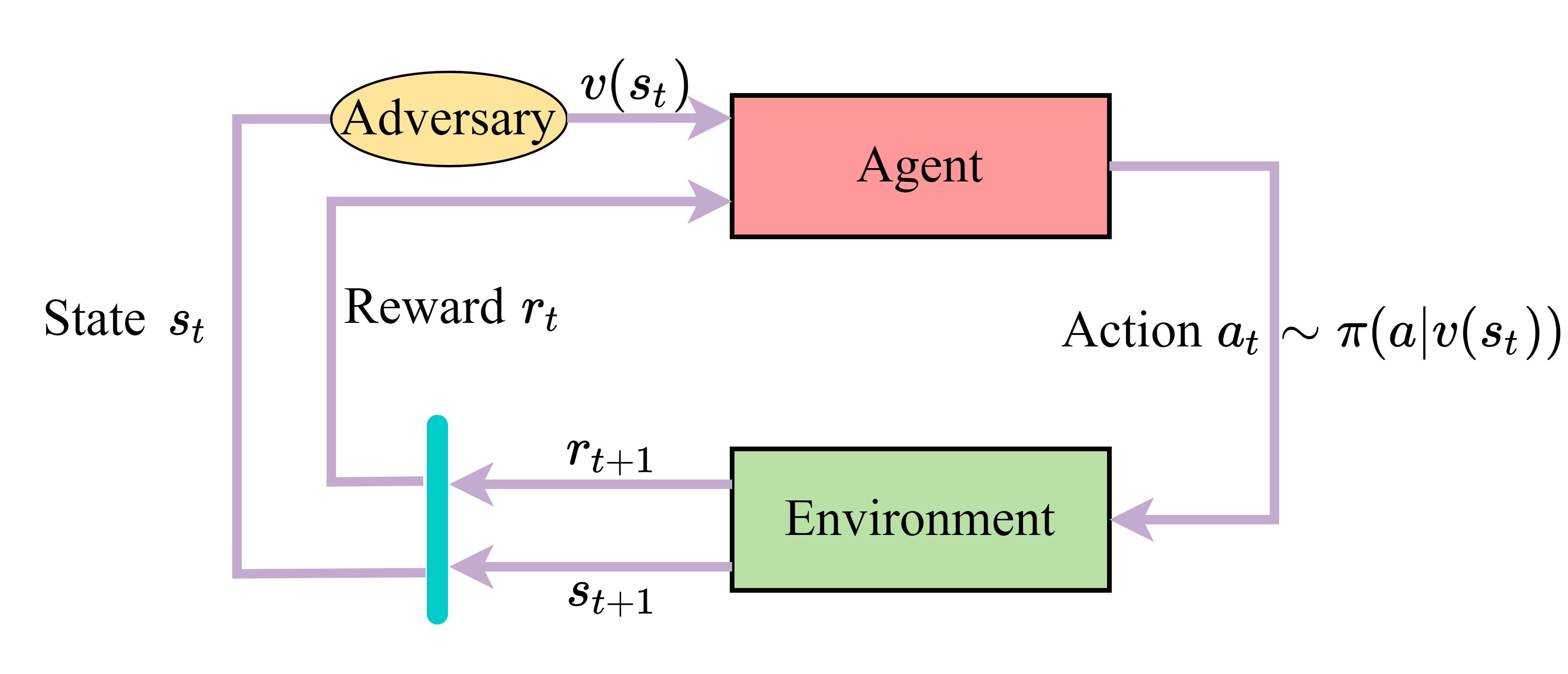}
    \caption{State-Adversarial Markov decision process.}
    \label{SA-MDP}
\end{figure}

Unlike a regular Markov decision process (MDP), the SA-MDP introduces an adversary $v(s)$ into the process. The adversary's aim is to disturb the agent's observation, causing the agent to take actions based on $\pi(a|v(s))$, while the environment still transitions from state $s$ rather than $v(s)$ to the next state. As a result, the agent's action from $\pi(a|v(s))$ may be sub-optimal, leading to a reduction in the agent's earned reward. In the context of IES scheduling, the adversary can be reflected as cyber-attacks. Specifically, the cyber-attacks we address may involve injecting false data or manipulating real-time data streams. These disturbances are represented as alterations in the input of state variables, impacting the decision-making process of scheduling algorithms, thereby introducing uncertainty and potential inaccuracies.The SA-MDP is shown in Fig. \ref{SA-MDP}.


\subsubsection{State}

The state is the environmental state space, the agent's perceived state at a particular time is denoted by ${{s}_{t}}\in S$. Here, the state is described as
{\setlength\abovedisplayskip{2pt}
\setlength\belowdisplayskip{2pt}
\begin{equation}\label{Eq47}
s_t=\left[\begin{array}{l}
S O C_t^{E S D}, S O C_t^{H S D}, \lambda_{\text {grid }, t}^P, \lambda_{\text {grid }, t}^Q, \\
P_{W T, t}, P_{\text {basic }, t}, Q_{\text {basic }, t}, H_{\text {basic }, t}, t
\end{array}\right]
\end{equation} }

\subsubsection{Action}

The action denotes the scheduling decision taken by IES. Here, the action at time $t$ can be described as follows:
{\setlength\abovedisplayskip{2pt}
\setlength\belowdisplayskip{2pt}
\begin{equation}\label{Eq48}
{{a}_{t}}=\left[ {{P}_{P2G,t}},{{P}_{EB,t}},{{Q}_{MT,t}},{\lambda_{rt, t}^H},{\lambda_{rt, t}^P},{\lambda_{rt, t}^Q} \right]
\end{equation} }

\subsubsection{Adversary}
The adversary represents realistic cyber-attack. We need to constrain the power of the adversary, i.e. ${{v}_{s}}\in B(s)$. In this study, the adversary is set to HLR attack, as described in (\ref{Eq25}).

\newtheorem{thm}{Theorem}
\begin{thm}
For a non-adversarial MDP with a policy $\pi$, and an optimal adversary $v$ in SA-MDP, the following holds for all states $s\in S$: 
{\setlength\abovedisplayskip{2pt}
\setlength\belowdisplayskip{2pt}
\begin{equation}\label{Eq49}
\max _{s \in \mathcal{S}}\left\{V^\pi(s)-\tilde{V}_v^\pi(s)\right\} \leq \alpha \max _{s \in \mathcal{S}} \max _{\hat{s} \in B(s)} \mathrm{D}_{T V}(\pi(\cdot \mid s), \pi(\cdot \mid \hat{s}))
\end{equation} }where ${{\text{D}}_{TV}}(\pi (\cdot \mid s),\pi (\cdot \mid \hat{s}))$ denotes the total variation distance between $\pi (\cdot \mid s)$and $\pi (\cdot \mid \hat{s})$; $V^\pi(s)$ and $\tilde{V}_v^\pi(s)$ represent the system performance under conditions without and with cyber-attack, respectively\cite{zhang2020robust}.
\end{thm}

Theorem 1 indicates that if ${{\text{D}}_{TV}}(\pi (\cdot \mid s),\pi (\cdot \mid \hat{s}))$ for any $\hat{s}$ within $B(s)$ is not too large, the performance difference between $\tilde{V}_{\nu }^{\pi }(s)$ in the SA-MDP environment and ${{V}^{\pi }}(s)$ in the MDP environment is bounded. So, we add a state adversarial regularizer to the policy network to generate a resilient scheduling strategy against cyber-attacks.

\subsubsection{Reward Function}
The reward function for maximizing the net profit of IES operator is augmented with a penalty term to maintain energy balance and accelerate the convergence, and is given by: 
{\setlength\abovedisplayskip{2pt}
\setlength\belowdisplayskip{2pt}
\begin{equation}\label{Eq50}
{{r}_{t}}({{s}_{t}},{{a}_{t}})=F({{s}_{t}},{{a}_{t}})-\left( C_{1,t}^{Penalty}({{s}_{t}},{{a}_{t}})+C_{2,t}^{Penalty}({{s}_{t}},{{a}_{t}}) \right)
\end{equation} }where $F$ is the net profit of the IES operator; $C_{1,t}^{Penalty}$ denotes the action change rate overrun penalty:
{\setlength\abovedisplayskip{2pt}
\setlength\belowdisplayskip{2pt}
\begin{equation}\label{Eq51}
C_{1, t}^{\text {Penaly }}=\sum_{a \in A}\left[\begin{array}{l}
\delta_{1, i} \max \left(a_{i, t}-a_{i, t-1}-a_{i, \mathrm{U} \max }, 0\right) \\
+\delta_{2, i} \max \left(a_{i, t-1}-a_{i, t}-a_{i, \mathrm{D} \max }, 0\right)
\end{array}\right]
\end{equation} }where ${{\delta }_{1,i}}$ and ${{\delta }_{2,i}}$ are the penalty coefficients of action change rate exceeding the upper and lower change limits. ${{a}_{i,\text{U}\max }}$ and ${{a}_{i,\text{D}\max }}$ are the upper and lower change limits of the action change rate. Here, $C_{2,t}^{Penalty}$ ensures supply-demand balance by preventing actions that could cause imbalance. 
{\setlength\abovedisplayskip{2pt}
\setlength\belowdisplayskip{2pt}
\begin{equation}\label{Eq52}
C_{2,t}^{Penalty}=\omega_t^H L R_t^H+\left(1-\omega_t^H\right) E L_t^H
\end{equation} }where $LR_{t}^{H}$ is the economic penalty on the demand-side of the heating system; $EL_{t}^{H}$ is the economic penalty on the supply-side of the heating system; $\omega_t^H$ is a binary variable, if $\omega_t^H=1$ means that the demand-side economic penalty is at this time, otherwise the supply-side economic penalty is at this time.

\vspace{-3mm}
\subsection{SA-SAC based Solution Process}

\begin{figure}[ht]
    \centering
    \includegraphics[width=3.6in,height=2.5in]{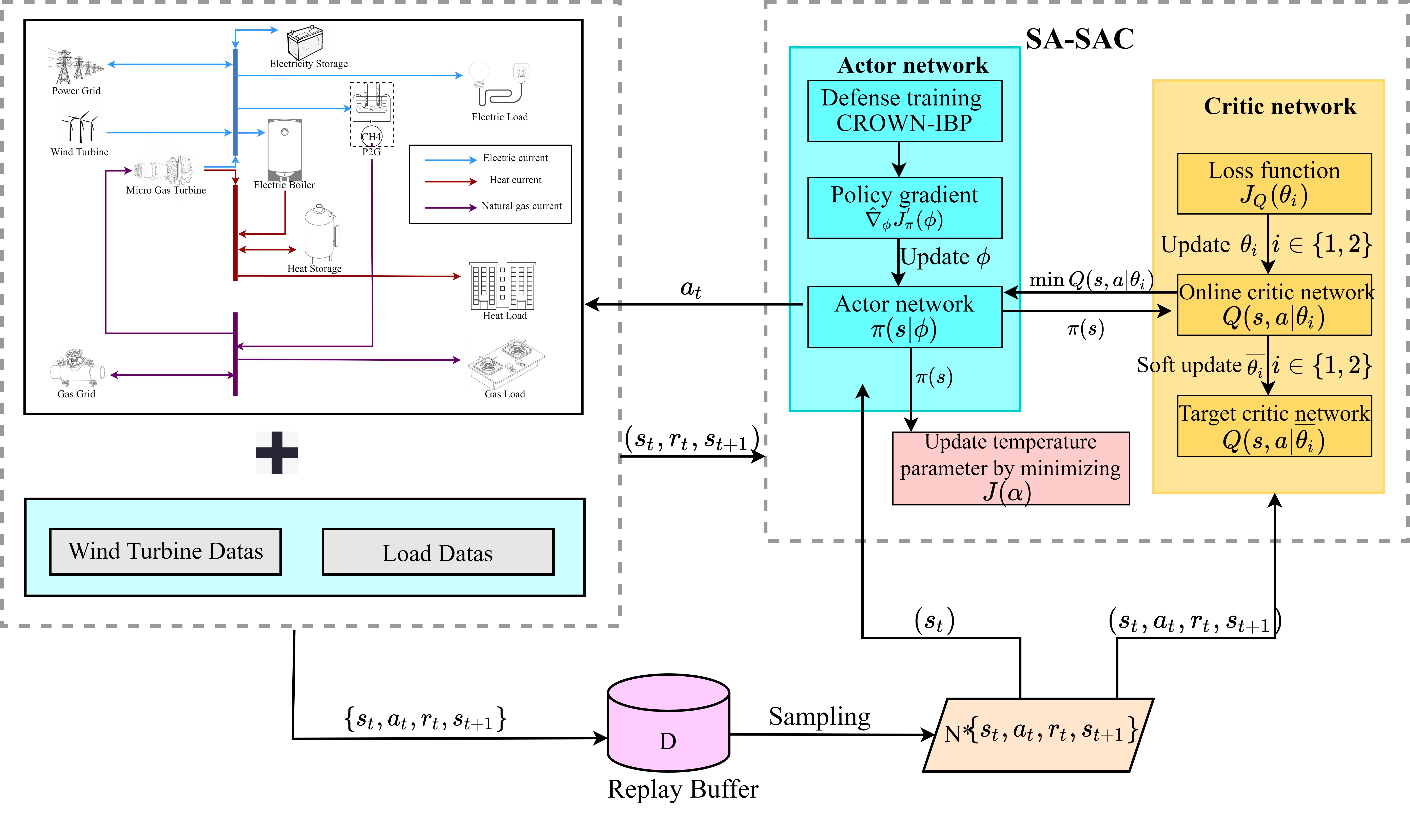}
    \caption{The IES scheduling framework based on the proposed SA-SAC.}
    \label{SA-SAC-graph}
\end{figure}

The proposed algorithm adopts a maximum entropy framework and trains the decision-making network using a certified defense training method to make its neural network robust to cyber-attacks perturbations. The scheduling framework based on the SA-SAC  is shown in Fig. \ref{SA-SAC-graph}. The objective function is 
{\setlength\abovedisplayskip{2pt}
\setlength\belowdisplayskip{2pt}
\begin{equation}\label{Eq57}
\pi^*=\arg \max _\pi \sum_t \mathbb{E}_{\left(s_t, \mathbf{a}_t\right) \sim \rho_\pi}\left[r\left(s_t, a_t\right)+\alpha \mathcal{H}\left(\pi\left(\cdot \mid s_t\right)\right)\right]
\end{equation} }where $\mathcal{H}$ is the entropy term of the action adopted in state ${{s}_{t}}$; $\alpha $ is the temperature parameter. The objective function contains two terms: the sum of cumulative rewards and the entropy term. This design motivates the agent to explore more action spaces while pursuing the maximum cumulative rewards. During the policy evaluation step of soft policy iteration, the modified Bellman equation is applied:

{\setlength\abovedisplayskip{2pt}
\setlength\belowdisplayskip{2pt}
\begin{equation}\label{Eq63}
\begin{aligned}
Q\left(s_t, a_t\right) & =r\left(s_t, a_t\right)+\gamma \mathbb{E}_{\mathbf{s}_{t+1} \sim p_\pi, \mathbf{a}_t \sim \pi}\left[Q\left(s_t, a_t\right)\right. \\
& \left.-\alpha \log \pi\left(a_t \mid s_t\right)\right]
\end{aligned}
\end{equation} }

The policy distribution is converted into a probability distribution in the form of energy function and is proportional to the exponential form of the Q function, at which point the action value function is called soft Q value.

For each state, the policy is updated using  KL divergence:
{\setlength\abovedisplayskip{2pt}
\setlength\belowdisplayskip{2pt}
\begin{equation}\label{Eq61}
\pi_{\text {new }}=\arg \min _{\pi^{\prime} \in \Pi} D_{K L}\left(\pi^{\prime}\left(\cdot \mid s_t\right) \| \frac{\exp \left(\frac{1}{\alpha} Q^{\pi_{\text {old }}}\left(s_t, \cdot\right)\right)}{Z^{\pi_{\text {old }}}\left(s_t\right)}\right)
\end{equation} }where $\Pi $ denotes the set of policy distribution; ${{Z}^{{{\pi }_{old}}}}({{s}_{t}})$ is the partition function that is used to normalize the distribution.

Consider a parameterized soft Q function $Q_\theta\left(s_t, a_t\right)$ and a tractable policy $\pi_\phi\left(a_t \mid s_t\right)$ with  $\theta$ and  $\phi$ as the parameters of their networks, respectively. The soft Q-function parameters are obtained by minimizing the soft Bellman residual:
{\setlength\abovedisplayskip{2pt}
\setlength\belowdisplayskip{2pt}
\begin{equation}\label{Eq62}
\begin{aligned}
J_Q(\theta) & =\mathbb{E}_{\left(s_t, a_t, s_{t+1}\right) \sim D, a_{t+1} \sim \pi_\phi}\left[1 / 2\left(Q_\theta\left(s_t, a_t\right)-r\left(s_t, a_t\right)\right.\right. \\
& \left.\left.-\gamma Q_{\bar{\theta}}\left(s_{t+1}, a_{t+1}\right)+\gamma \alpha \log \pi_\phi\left(a_{t+1} \mid s_{t+1}\right)\right)^2\right]
\end{aligned}
\end{equation} }
where $\overline{\theta }$ is the parameters of the target soft Q-function, and soft update is used for the parameters update of the target soft Q-function in the training, i.e., $\overline{\theta }\leftarrow \tau \theta +(1-\tau )\overline{\theta }$.

Here, we use two Q-functions and select the one with the lower value during each iteration. For the update of the policy network, the policy parameter can be learned by 
{\setlength\abovedisplayskip{2pt}
\setlength\belowdisplayskip{2pt}
\begin{equation}\label{Eq64}
J_\pi(\phi)=\mathbb{E}_{s_t \sim D, a_t \sim \pi_\phi}\left[\alpha \log \left(\pi_\phi\left(a_t \mid s_t\right))-Q_\theta\left(s_t, a_t\right)\right]\right.
\end{equation} }

In addition, we introduce a reparameterization trick ${{f}_{_{\phi }}}({{\varepsilon }_{t}};{{s}_{t}})$, where $\varepsilon $ is the action noise sampled from the standard normal distribution. Based on Theorem 1, to obtain the resilient scheduling strategy under attack, we need to find ${{\max }_{\hat{s}\in B(s)}}{{\text{D}}_{TV}}(\pi (\cdot \mid s),\pi (\cdot \mid \hat{s}))$. To establish an upper bound for the total variation distance of the stochastic policy distribution, we employ the KL divergence as follows:
{\setlength\abovedisplayskip{2pt}
\setlength\belowdisplayskip{2pt}
\begin{equation}\label{Eq65}
\mathrm{D}_{\mathrm{TV}}(\pi(a \mid s), \pi(a \mid \hat{s})) \leq \sqrt{\frac{1}{2} \mathrm{D}_{\mathrm{KL}}(\pi(a \mid s) \| \pi(a \mid \hat{s}))}
\end{equation} }

By ignoring constant terms, regularizing the KL divergence over all $s$ from sampled trajectories and all $\hat{s} \in B(s)$ results in the state-adversarial regularizer:
{\setlength\abovedisplayskip{2pt}
\setlength\belowdisplayskip{2pt}
\begin{equation}\label{KL}
\begin{aligned}
\mathcal{R}_{S A-S A C}\left(\phi_\mu\right) & =\sum_s \max _{\hat{s} \in B(s)}\left(\pi_\phi^\mu(\hat{s})-\pi_\phi^\mu(s)\right)^{\top} \Sigma^{-1}\left(\pi_\phi^\mu(\hat{s})\right. \\
\left.-\pi_\phi^\mu(s)\right) & =\sum_s \max _{\hat{s} \in B(s)}\left\|\pi_\phi^\mu(\hat{s})-\pi_\phi^\mu(s)\right\|_2^2 \\
& =\sum_s \max _{\hat{s} \in B(s)} \mathcal{R}_s\left(\hat{s}, \phi_\mu\right)
\end{aligned}
\end{equation} }

The defense training strategy, which integrates the rapid interval bound propagation (IBP) with CROWN's accurate convex relaxation approach, to accelerate training while preventing excessive regularization of the network. The incorporation of CROWN-IBP into the DRL framework augments the resilience of the policy network, guaranteeing dependable decision-making and optimal outputs for the IES amidst conditions of cyber-attack. CROWN-IBP is used to derive linear upper and lower bounds of $\pi_\phi^\mu(\hat{s})$:

{\setlength\abovedisplayskip{2pt}
\setlength\belowdisplayskip{2pt}
\begin{equation}\label{Eq66}
l_{\phi_\mu}(s) \leq \pi_\phi^\mu(\hat{s}) \leq u_{\phi_\mu}(s) \quad 
\end{equation} }where the bounds $u_{\phi_\mu}$ and $l_{\phi_\mu}$ are calculated as follows:
{\setlength\abovedisplayskip{2pt}
\setlength\belowdisplayskip{2pt}
\begin{equation}\label{Eq_defend}
\left\{\begin{array}{l}
u_{\phi_\mu}(s)=(1-\beta) \overline{m}_{\mathrm{IBP}}(s, \epsilon)+\beta \overline{m}_{\mathrm{CROWN}}(s, \epsilon) \\
l_{\phi_\mu}(s)=(1-\beta) \underline{m}_{\mathrm{IBP}}(s, \epsilon)+\beta \underline{m}_{\mathrm{CROWN}}(s, \epsilon)
\end{array}\right.
\end{equation} }where $\epsilon$ refers to perturbations caused by cyber-attacks; ${{m}_{\text{IBP}}}(s,\epsilon )$ and ${{m}_{\text{CROWN}}}(s,\epsilon )$ are the tight linear boundaries generated by IBP and CROWN, respectively; the hyper-parameter $\beta$ in CROWN-IBP is scheduled as suggested in \cite{zhang2019towards}.

We employed CROWN-IBP to calculate interval bounds for attack perturbations on each state, ensuring the model's generalizability across various attack scenarios. Regularization coefficients within the adversarial regularizer further mitigate overfitting, enabling our model to adapt to evolving threats and learn generalized resilience strategies effectively.

Thus, we can calculate the upper bound $\overline{\mathcal{R}}_s\left(\phi_\mu\right)$ of $\mathcal{R}_s\left(\hat{s}, \phi_\mu\right)$ and the upper limit of the distance as follows:
{\setlength\abovedisplayskip{2pt}
\setlength\belowdisplayskip{2pt}
\begin{equation}\label{Eq67}
L_{S A-S A C}(\phi)=\mathbb{E}_{i \sim N}\left\|u_{\phi_\mu}\left(s_i\right)-l_{\phi_\mu}\left(s_i\right)\right\|_2^2=\mathbb{E}_{i \sim N} \overline{\mathcal{R}}_i\left(\phi_\mu\right)
\end{equation} }

Then we add the upper limit $L_{S A-S A C}(\phi)$ of the distance to the policy gradient loss to obtain the approximate gradient of $J_{\pi }^{'}(\phi )$, i.e.,

{\setlength\abovedisplayskip{2pt}
\setlength\belowdisplayskip{2pt}
\begin{equation}\label{Eq68}
\begin{gathered}
\hat{\nabla}_\phi J_\pi^{\prime}(\phi)=\nabla_\phi \alpha \log \left(\pi_\phi\left(a_t \mid s_t\right)\right)+\left(\nabla_{a_t} \alpha \log \left(\pi_\phi\left(a_t \mid s_t\right)\right)\right. \\
\left.-\nabla_{a_t} Q\left(s_{t+1}, a_{t+1}\right)\right) \nabla_\phi f_\phi\left(\varepsilon_t ; s_t\right)+\kappa \nabla_\phi L_{S A-S A C}(\phi)
\end{gathered}
\end{equation} }where ${{a}_{t}}$ is evaluated at ${{f}_{_{\phi }}}({{\varepsilon }_{t}};{{s}_{t}})$, i.e., ${{a}_{t}}={{f}_{_{\phi }}}({{\varepsilon }_{t}};{{s}_{t}})=\pi _{\phi }^{\mu }({{s}_{t}})+{{\varepsilon }_{t}}\cdot \pi _{\phi }^{\sigma }({{s}_{t}})$; $\kappa $ is the regularization term coefficient.

In addition, for the temperature parameter $\alpha$, we introduce an adaptive update in the form of minimizing $J(\alpha )$, i.e.,
{\setlength\abovedisplayskip{2pt}
\setlength\belowdisplayskip{2pt}
\begin{equation}\label{Eq69}
J(\alpha)=\mathbb{E}_{a_t, \pi_t}\left[-\alpha \log \left(\pi_\phi\left(a_t \mid s_t\right)-\alpha H_0\right]\right.
\end{equation} }where ${{H}_{0}}$ is set to $-\dim(a)$, the dimension of the action. 

To encapsulate the preceding steps, the detailed solution process based on the SA-SAC is outlined in Algorithm \ref{alg:code}.
\begin{algorithm}[!t]\label{code} 
\small
\renewcommand{\algorithmicrequire}{\textbf{Initialize:}}
\renewcommand{\algorithmicensure}{\textbf{return}}
    \caption{State-Adversarial Soft Actor-Critic (SA-SAC)} 
    \label{alg:code}
	\begin{algorithmic}[1] 
	\REQUIRE the critic network and actor network with parameters ${{\theta }_{1}}$, ${{\theta }_{2}}$ and $\phi $
	\REQUIRE the target network with weights $\overline{{{\theta }_{1}}}\leftarrow {{\theta }_{1}}$ and $\overline{{{\theta }_{2}}}\leftarrow {{\theta }_{2}}$
    \REQUIRE the replay pool capacity D
    \FOR{each iteration}
    \FOR{each environment step} 
    \STATE Sample action from the policy ${{a}_{t}}\sim{{\pi }_{\phi }}({{a}_{t}}|{{s}_{t}})$
    \STATE Sample state transition from the environment $s_{t+1} \sim p\left(s_{t+1} \mid s_t, a_t\right)$
    \STATE Store state transition and reward in the replay pool$D\leftarrow D\bigcup \left\{ ({{s}_{t}},{{a}_{t}},r({{s}_{t}},{{a}_{t}}),{{s}_{t+1}}) \right\}$
    \ENDFOR
    \FOR{each gradient step}
    \STATE Update the Q-function ${{J}_{Q}}({{\theta }_{i}})$ parameters, ${{\theta }_{i}}\leftarrow {{\theta }_{i}}-{{\lambda }_{Q}}{{\hat{\nabla }}_{{{\theta }_{i}}}}{{J}_{Q}}({{\theta }_{i}})$for $i\in \left\{ 1,2 \right\}$
    \STATE Obtain upper and lower bounds on ${{\pi }_{\phi }}\left( {{s}_{n}} \right)$ using CROWN-IBP: $l_{\phi_\mu}(s) \leq \pi_\phi^\mu(\hat{s}) \leq u_{\phi_\mu}(s)$ for $s\in B({{s}_{n}})$, $n\in \left[ N \right]$
    \STATE Upper bound on distance: $L_{S A-S A C}(\phi)=\mathbb{E}_{i \sim N} \overline{\mathcal{R}}_i\left(\phi_\mu\right)$
    \STATE Update policy $J_{\pi }^{'}(\phi )={{J}_{\pi }}(\phi )+\kappa {{L}_{SA}}(\phi )$ weight, $\phi \leftarrow \phi -{{\lambda }_{\pi }}{{\hat{\nabla }}_{\phi }}J_{\pi }^{'}(\phi )$
    \STATE Update temperature parameter, $\alpha \leftarrow \alpha -\lambda {{\hat{\nabla }}_{\alpha }}J(\alpha )$
    \STATE Soft update target network weights, $\overline{{{\theta }_{i}}}\leftarrow \tau {{\theta }_{i}}+(1-\tau )\overline{{{\theta }_{i}}}$for $i\in \left\{ 1,2 \right\}$
    \ENDFOR
    \ENDFOR
    \STATE Output policy network as the resilient scheduling model of integrated energy system
  \end{algorithmic}
\end{algorithm}

\section{Case Study}
To validate the efficacy of our approach, we perform a case study on the IES illustrated in Fig. \ref{CIES}. Our proposed method is developed using Python 3.8 and implemented with PyTorch 1.6.0. The simulations are conducted on a computer equipped with an AMD R7-5800H CPU and 16 GB of RAM.

The parameters of the SA-SAC are given as follows: The actor and critic use ReLU as the activation function of all hidden layers, and the Adam optimizer is used to update the network weights. The learning rates of actor network and critic network are 0.0005 and 0.002, respectively. The discount factor is set to 0.95, the mini-batch size is 256, the target network is updated using a soft update rate of 0.005, and the training epoch is set as 1000. In addition, we train the actor network without actor regularizer for the first 500 episodes, then gradually increase the perturbation $\epsilon$ from the cyber-attack on the actor neural network in 300 episodes, eventually maintaining the target $\epsilon$ during 200 episodes of training.

\subsection{Settings in Test Case}

The length of a scheduling cycle is set to 24 hours, with a one-hour interval for each time step. The main parameters of the devices are shown in Table \ref{tab2}.

\begin{table}[ht]   
  \centering
  \caption{Main parameters of IES}\label{tab2}
    \begin{tabular}{cccc}  
    \specialrule{0.5pt}{0.5pt}{0.5pt}
    \specialrule{0.5pt}{0.5pt}{0.5pt}
    Parameter & Value & Parameter & Value \\
    \midrule
    ${{H}_{EB,\max }}(\operatorname{kW})$ & 300 &  ${\varsigma _{t}^{TSE}}$ & $[-2,2]$ \\
    ${{\eta }^{EB}}$    & 0.99  & $\eta $     & $\operatorname{N}(10,6)$           \\ 
    $Q_{\max }^{grid}({{\operatorname{m}}^{3}})$  & 80  & ${{\sigma }_{t}}(\operatorname{kW})$  & $\operatorname{N}(0.2,0.2)$   \\ 
    ${{Q}_{MT,\min }}({{\operatorname{m}}^{3}})$  & 10 & $P_{\max }^{grid}\text{(kW)}$ & 1000  \\       
    ${{Q}_{MT,\max }}({{\operatorname{m}}^{3}})$ & 40  & ${{P}_{P2G,\min }}(\operatorname{kW})$       & 100                                               \\
    $\Delta {{Q}_{MT,\min }}({{\operatorname{m}}^{3}})$  & -10  & ${{P}_{P2G,\max }}(\operatorname{kW})$ & 500                                                     \\
    $\Delta {{Q}_{MT,\max }}({{\operatorname{m}}^{3}})$  & 10 & $\Delta {{P}_{P2G,\min }}(\operatorname{kW})$ & -200                                                    \\
    $\eta _{e}^{MT}$  & 0.4                    & $\Delta {{P}_{P2G,\max }}(\operatorname{kW})$ & 200                                                     \\
    $\eta _{h}^{MT}$ & 0.5  & ${{\eta }_{P2G}}$ & 0.6                              \\
    $\eta _{loss}^{MT}$  & 0.1   &  $C_{\max }^{HSD}\text{(kWh)}$ & 180            \\
    $C_{\max }^{ESD}\text{(kWh)}$ & 200 & $P_{ch,\max }^{HSD},P_{dc,\max }^{HSD}\text{(kW)}$ & 90                                                                          \\
    $P_{ch,\max }^{ESD},P_{dc,\max }^{ESD}\text{(kW)}$ & 100 & $\eta _{ch}^{HSD},\eta _{dc}^{HSD}$ & 1                                                               \\
    $\eta _{ch}^{ESD},\eta _{dc}^{ESD}$ & 0.9 & ${\varsigma _{t}^{TSQ}}$ & $[-2,2]$                                                                         \\
    $K\text{(W}\cdot {{\text{m}}^{\text{-2}}}\text{)}$ & 0.5 & ${\varsigma _{t}^{HC}}$ & $[0,1]$                                       \\
    ${{c}_{air}}\text{(kJ}\cdot \text{k}{{\text{g}}^{\text{-1}}}\cdot {}^\circ {{C}^{\text{-1}}}\text{)}$ & 1.007 & $F\text{(}{{\text{m}}^{\text{2}}}\text{)}$ & 2400                                        \\
    ${{\rho }_{air}}\text{(kg}\cdot {{\text{m}}^{\text{-3}}}\text{)}$ & 1.2 & $V\text{(}{{\text{m}}^{\text{3}}}\text{)}$ & 36000                                        \\
    \specialrule{0.5pt}{0.5pt}{0.5pt}
    \specialrule{0.5pt}{0.5pt}{0.5pt}
    \end{tabular}%
  \label{tab:addlabel2}%
\end{table}%

\subsection{Analysis of Algorithm Robustness}
\begin{figure}[htbp]
    \centering
    \includegraphics[width=2.3in,height=1.5in]{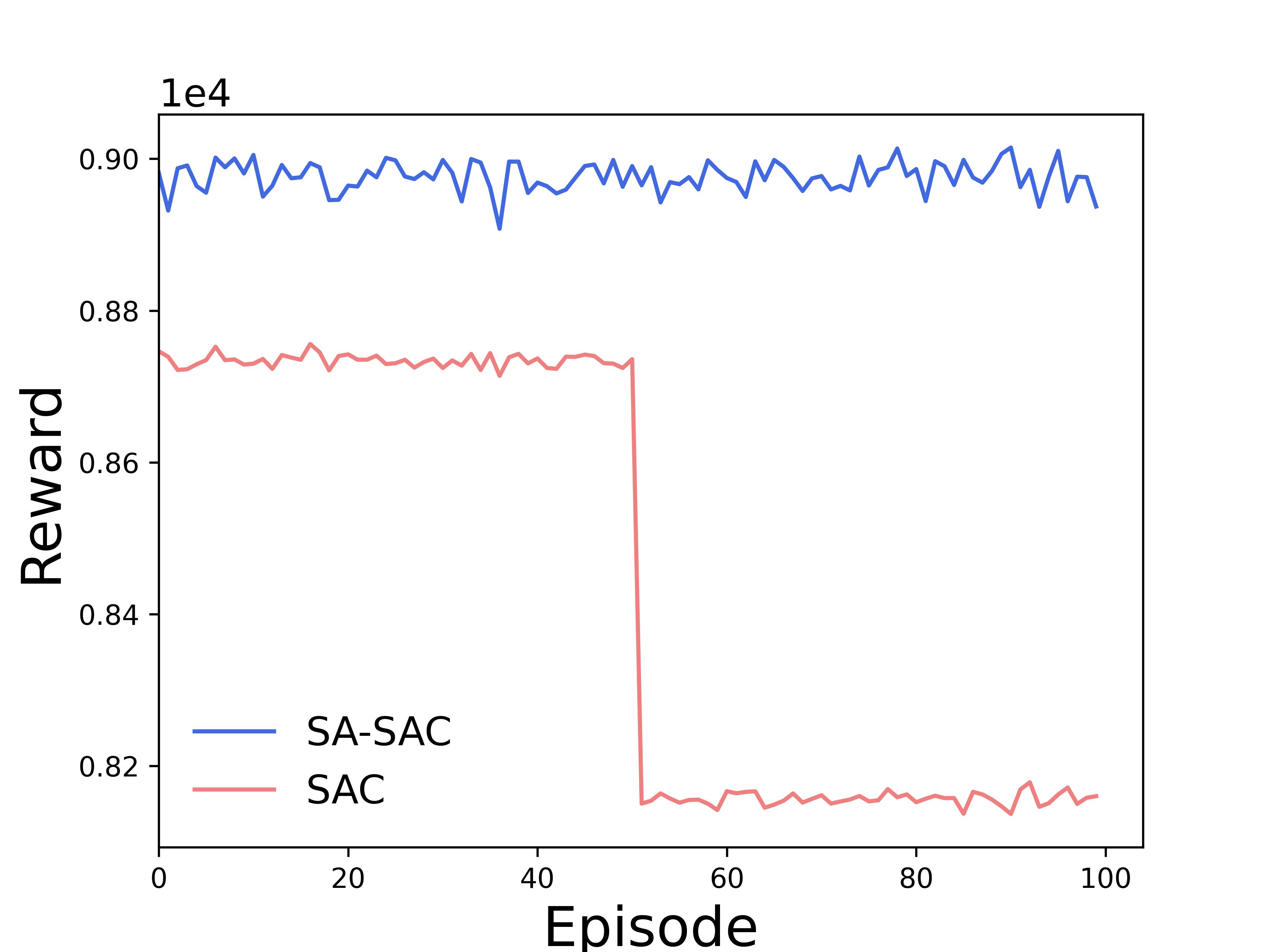}
    \caption{Robustness comparison of the SA-SAC and SAC to cyber-attack.}
    \label{Algorithm_attack}
\end{figure}

To assess the robustness of our decision network under cyber-attack scenarios, we compared the performance of SAC and our proposed algorithms in both regular MDP and SA-MDP environments. The SA-MDP environment's new test data set was created by simulating cyber-attacks on the IES. Fig. \ref{Algorithm_attack} illustrates this comparison, with the first 50 episodes in a non-attacked MDP environment and the last 50 in an HLR-attacked SA-MDP environment.

The proposed method, as illustrated in Fig. \ref{Algorithm_attack}, achieves higher rewards post-attack compared to SAC, demonstrating its resilience in scheduling under cyber-attacks. This success is due to the enhanced robustness of its decision-making network, reinforced by CROWN-IBP, which leads to effective scheduling decisions without losses. In contrast, the untrained SAC algorithm shows marked vulnerability to malicious inputs, significantly reducing reward value.

\subsection{Analysis of Scheduling Results Using SA-SAC}

\begin{figure}[htbp]
\centering  
\subfigure[]{   
\begin{minipage}[t]{0.49\linewidth}\centering    
\includegraphics[height=3.3cm,width=4.2cm]{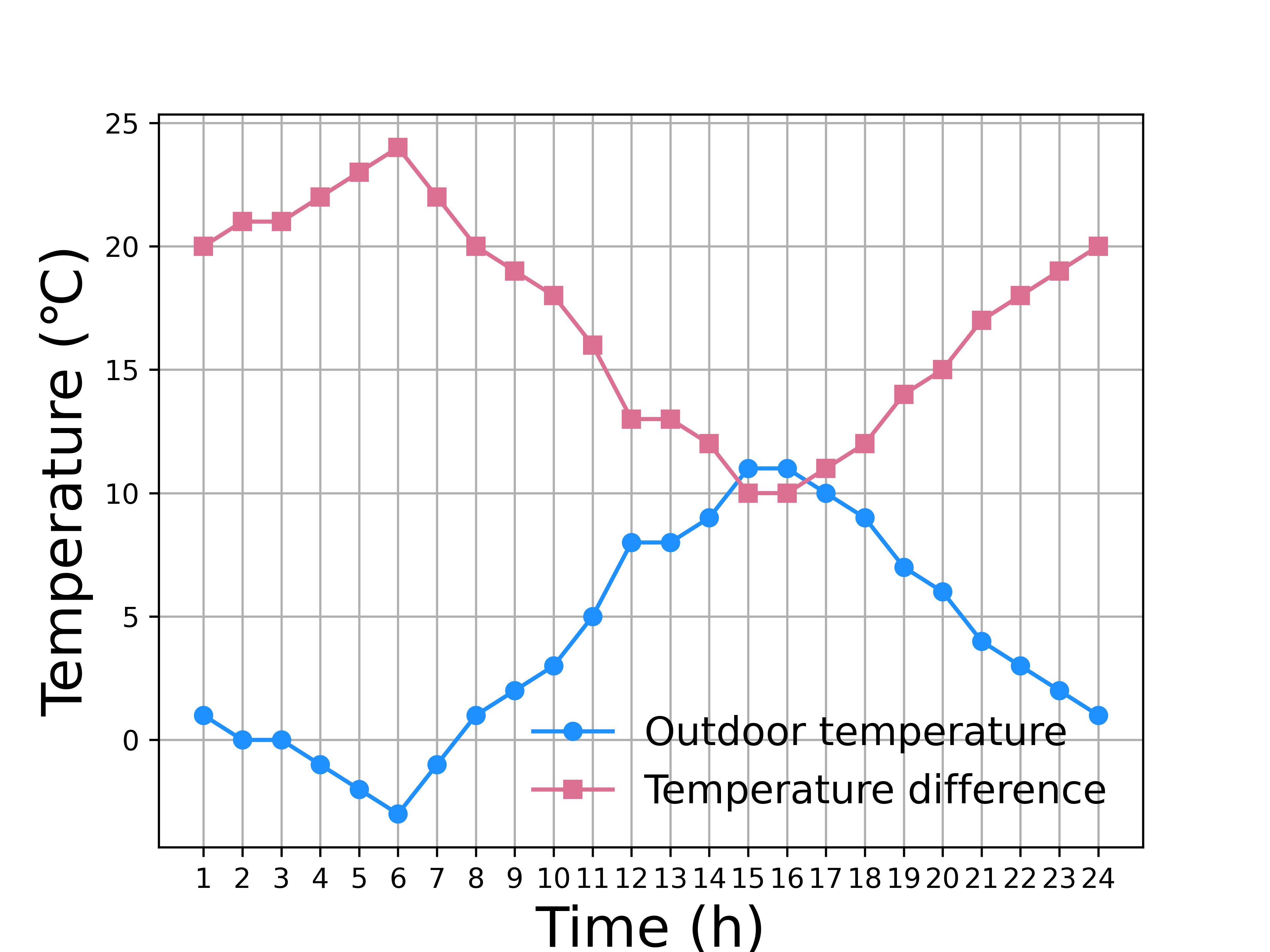}\end{minipage}}\subfigure[]{ 
\begin{minipage}[t]{0.49\linewidth}
\centering    
\includegraphics[height=3.3cm,width=4.2cm]{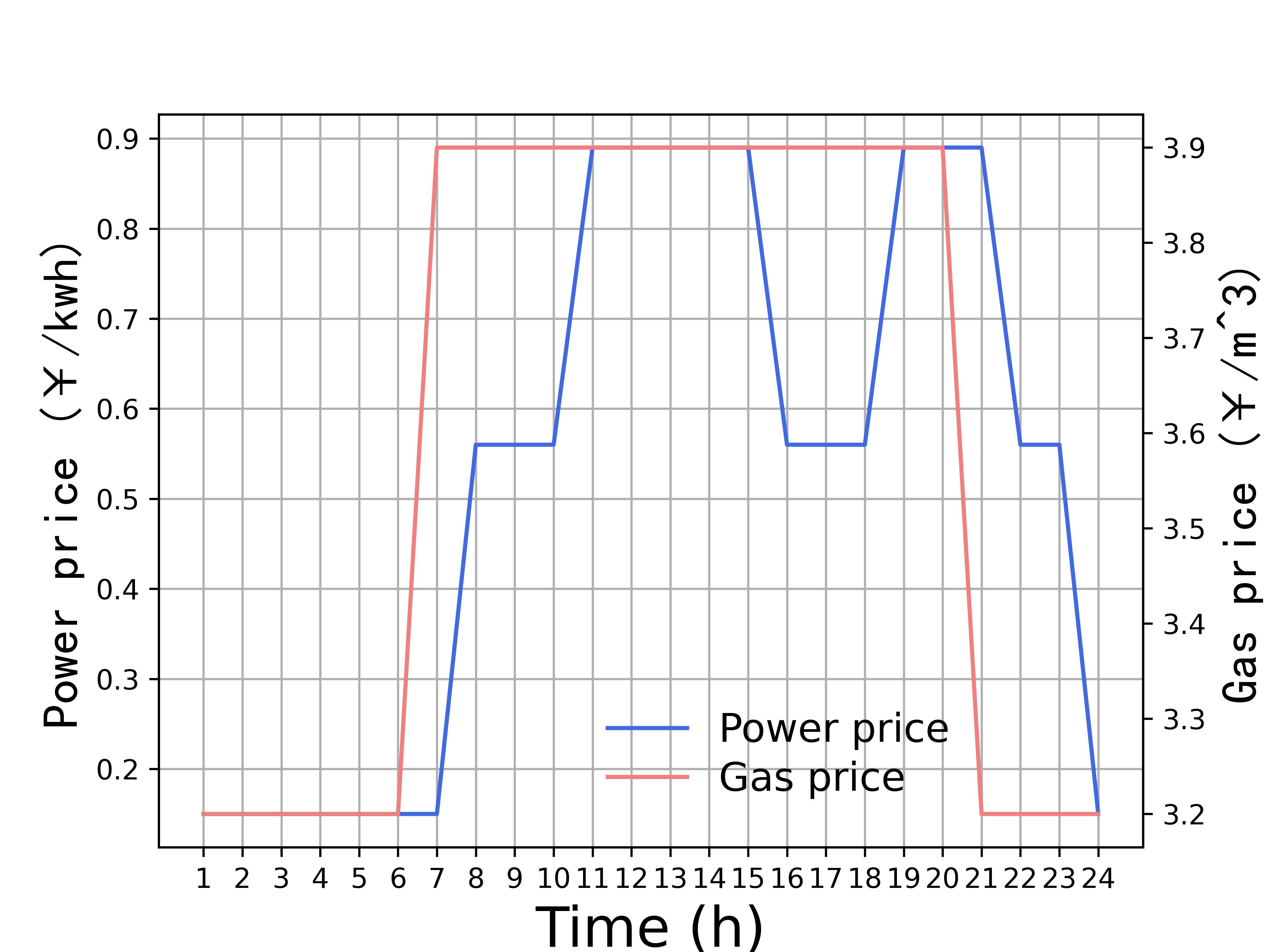}
\end{minipage}
}
\caption{Daily temperature curve and energy price.}    
\label{temperature and energy price}    
\end{figure}

\begin{figure}[htbp]
\begin{minipage}[t]{0.49\linewidth}
\centering
\includegraphics[height=3.3cm,width=4.2cm]{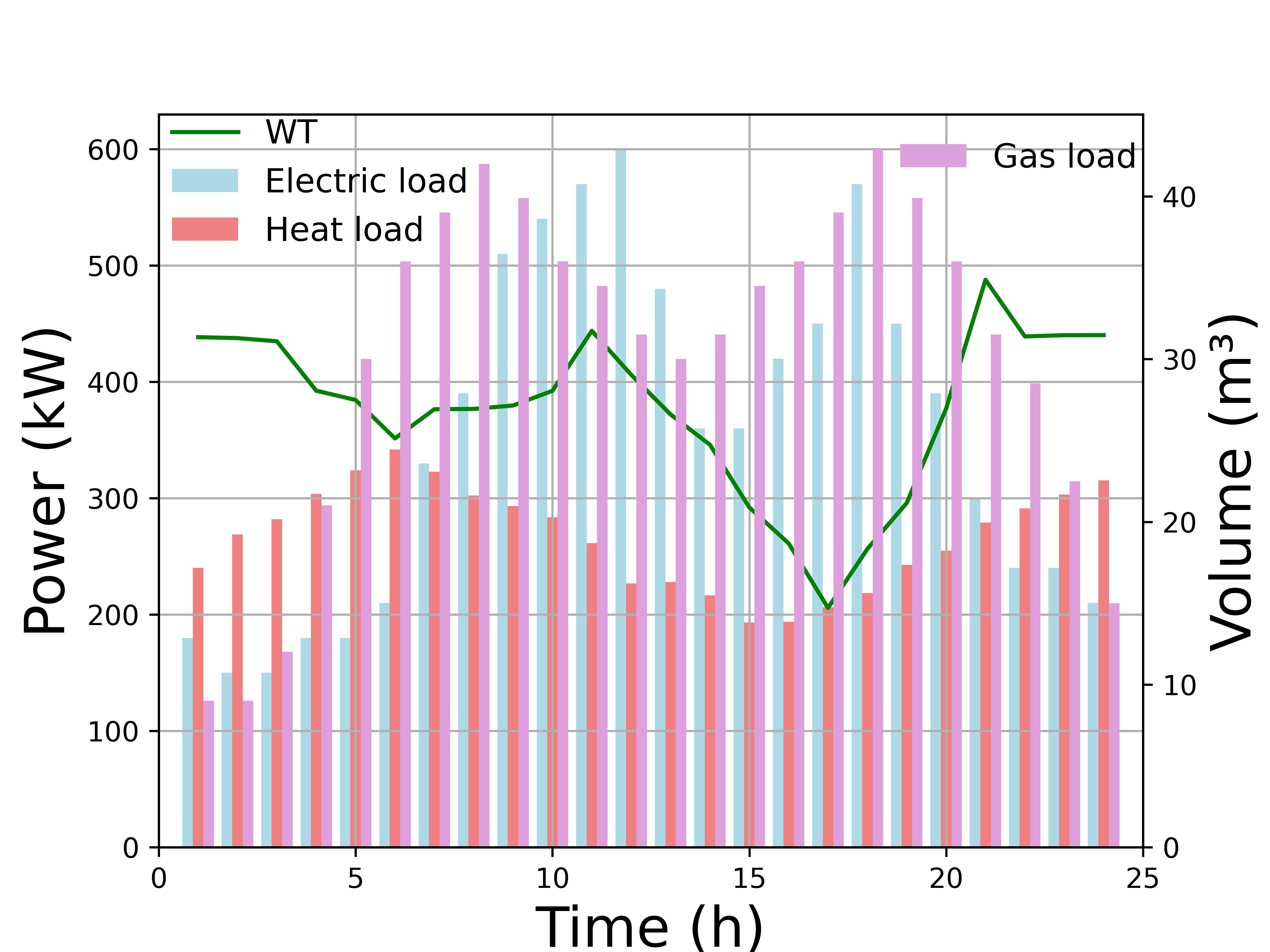}
\caption{WT outputs and daily load variations.}
\label{WT}
\end{minipage}%
\hfill
\begin{minipage}[t]{0.49\linewidth}
\centering
\includegraphics[height=3.3cm,width=4.2cm]{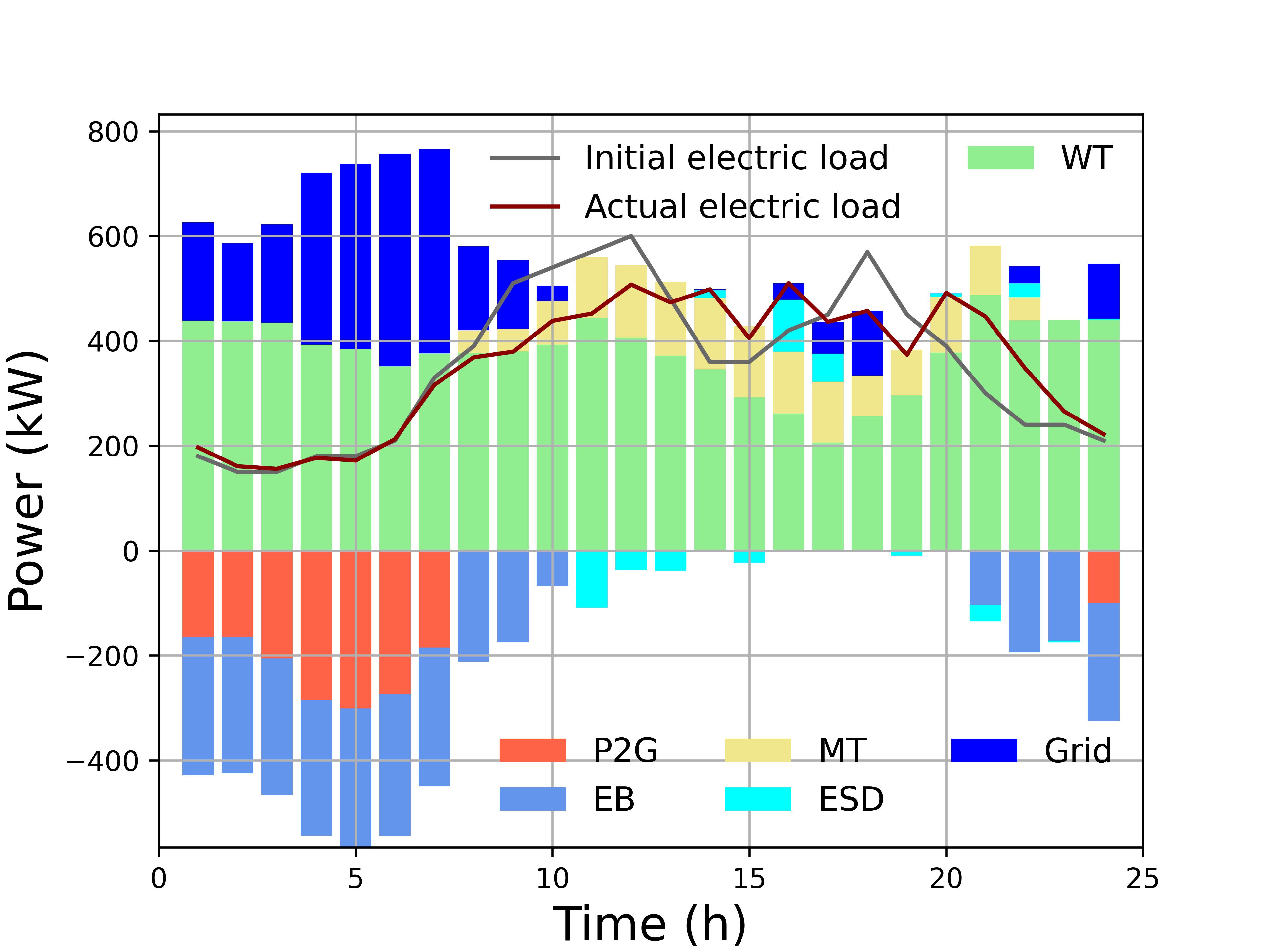}
\caption{Energy dispatching scheme for electrical subsystem.}
\label{ele_scheduling}
\end{minipage}
\end{figure}

To validate our approach, we utilize typical daily energy demand and wind turbine data as the foundational data for subsequent analysis. Fig. \ref{temperature and energy price}(a) displays the daily temperature curve and indoor-outdoor temperature difference, while Fig. \ref{temperature and energy price}(b) illustrates electricity and gas purchasing prices, with selling prices being half of the purchasing prices. Fig. \ref{WT} shows the load and WT outputs.

\begin{figure}[htbp]
\begin{minipage}[t]{0.49\linewidth}
\centering
\includegraphics[height=3.3cm,width=4.2cm]{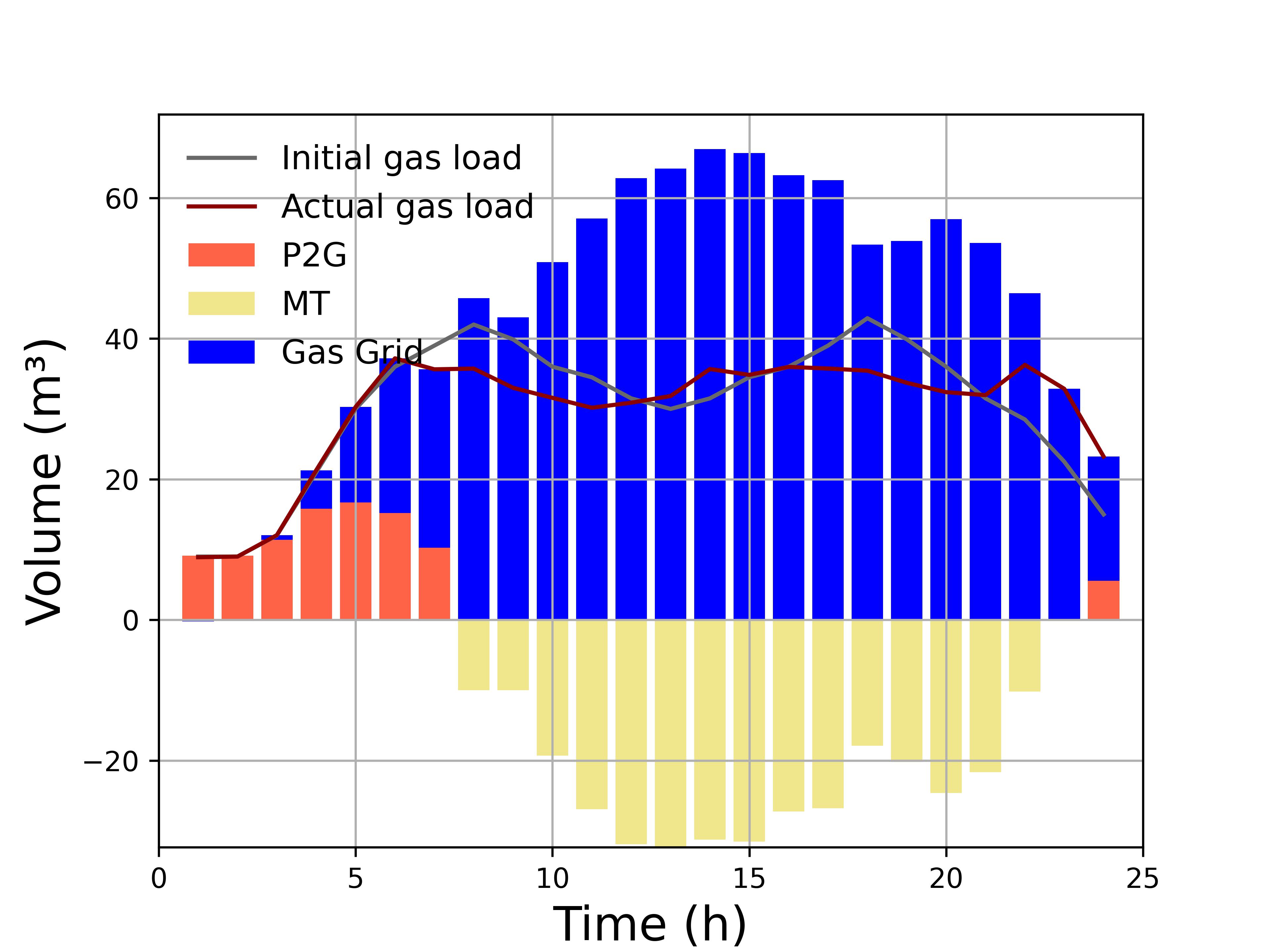}
\caption{Energy dispatching scheme for natural gas subsystem.}
\label{gas_scheduling}
\end{minipage}%
\hfill
\begin{minipage}[t]{0.49\linewidth}
\centering
\includegraphics[height=3.3cm,width=4.2cm]{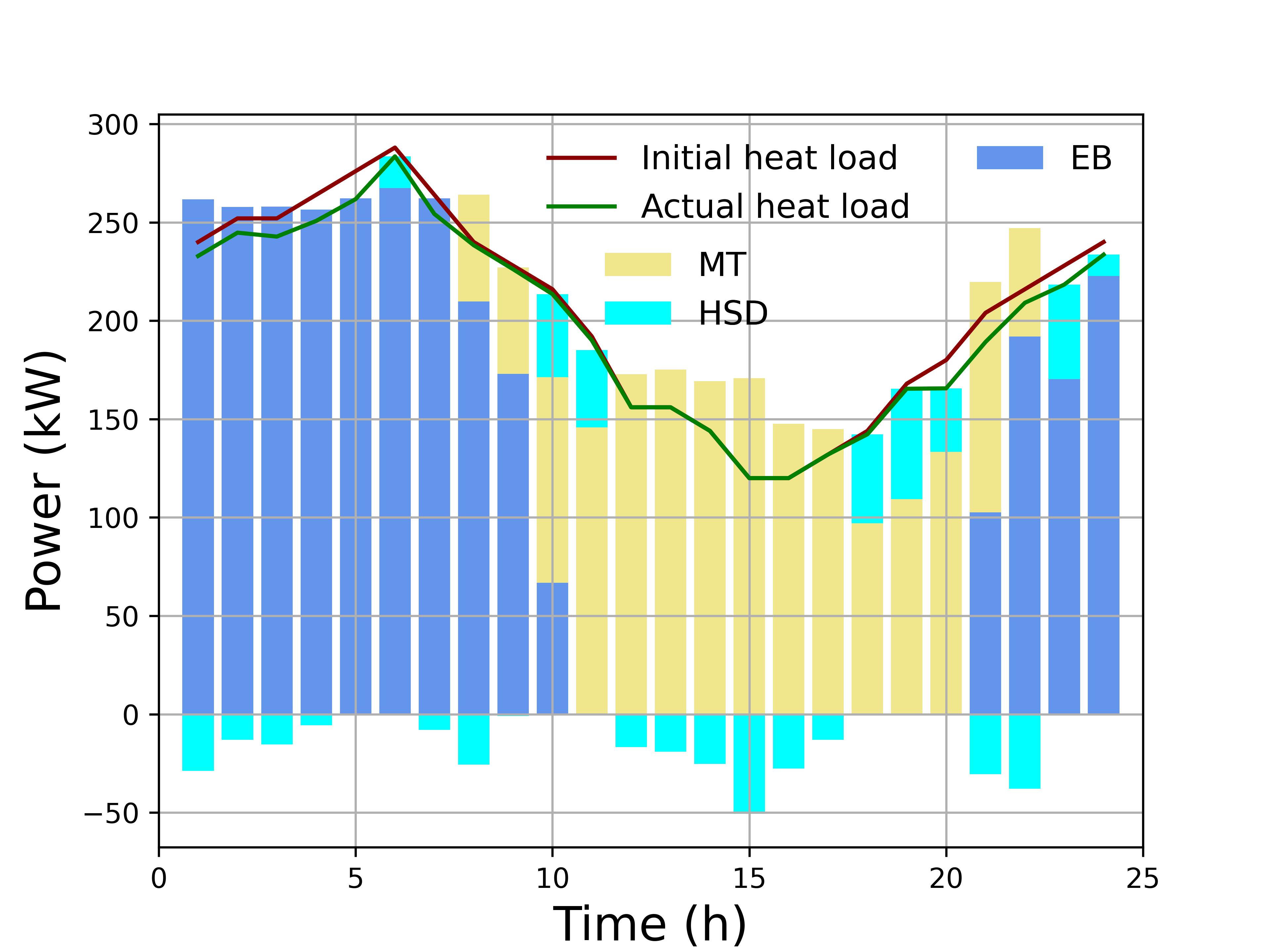}
\caption{Energy dispatching scheme for heating subsystem.}
\label{heat_scheduling}
\end{minipage}
\end{figure}

Upon completion of the training process, the policy network is employed for IES scheduling. The energy dispatching schemes for the electrical, natural gas, and heating subsystems are shown in Fig. \ref{ele_scheduling}, Fig. \ref{gas_scheduling}, and Fig. \ref{heat_scheduling}, respectively.

\subsubsection{Scheduling Schemes Analysis for Each Device}
Figs. \ref{ele_scheduling}-\ref{heat_scheduling} demonstrate that the decision network's scheduling strategy meets the needs of each energy subsystem. Fig. \ref{ele_scheduling} shows priority given to WT output for renewable energy use, grid electricity use mainly during 0:00-7:00, and increasing MT usage with higher loads and prices. Fig. \ref{gas_scheduling} indicates P2G energy conversion mainly during off-peak hours and MT operation during peak hours, illustrating improved system flexibility through P2G and MT integration. Fig. \ref{heat_scheduling} reveals that MT meets daytime heat demands, while at night, EB caters to increased heat needs due to lower temperatures and electricity prices, and higher WT output.

\subsubsection{Flexible Demand Response Analysis}
\begin{figure}[htbp]
    \centering
    \includegraphics[width=2.3in,height=1.5in]{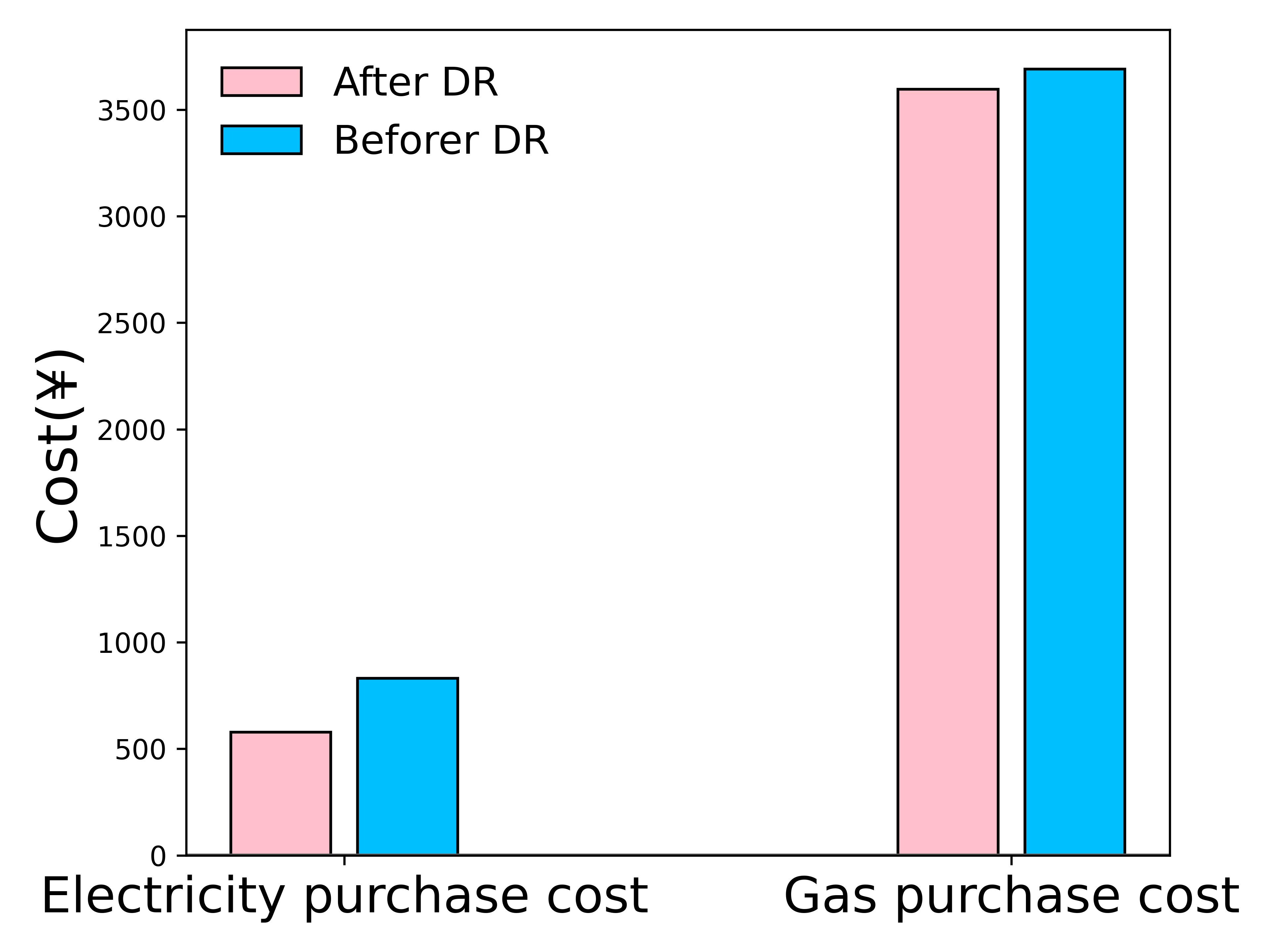}
    \caption{Comparison of the cost of energy purchased by IES from external energy supply networks before and after demand response.}
    \label{DR_cost}
\end{figure}

Figs. \ref{ele_scheduling} and \ref{gas_scheduling} show that electric and natural gas loads align with peak and valley prices, demonstrating a "peak shaving and valley filling" pattern post-demand response. Fig. \ref{heat_scheduling} indicates a notable thermal load decrease during nighttime due to reduced temperature perception and wider thermal comfort ranges. Additionally, Fig. \ref{DR_cost} compares the amount of electricity and natural gas purchased from external networks before and after demand response implementation, highlighting its economic impact on the IES.

As shown in Fig \ref{DR_cost}, the introduction of demand response has led to a rational allocation of energy use within the IES through time-shifted electrical load and time-shifted gas load. This has resulted in a reduction in the system's electricity and natural gas procurement costs, as well as an increase in the utilization of renewable energy.

\vspace{-4mm}
\subsection{Impact Analysis of Cyber-Attack on Scheduling}

\begin{figure}[htbp]
\begin{minipage}[t]{0.49\linewidth}
\centering
\includegraphics[height=3.3cm,width=4.2cm]{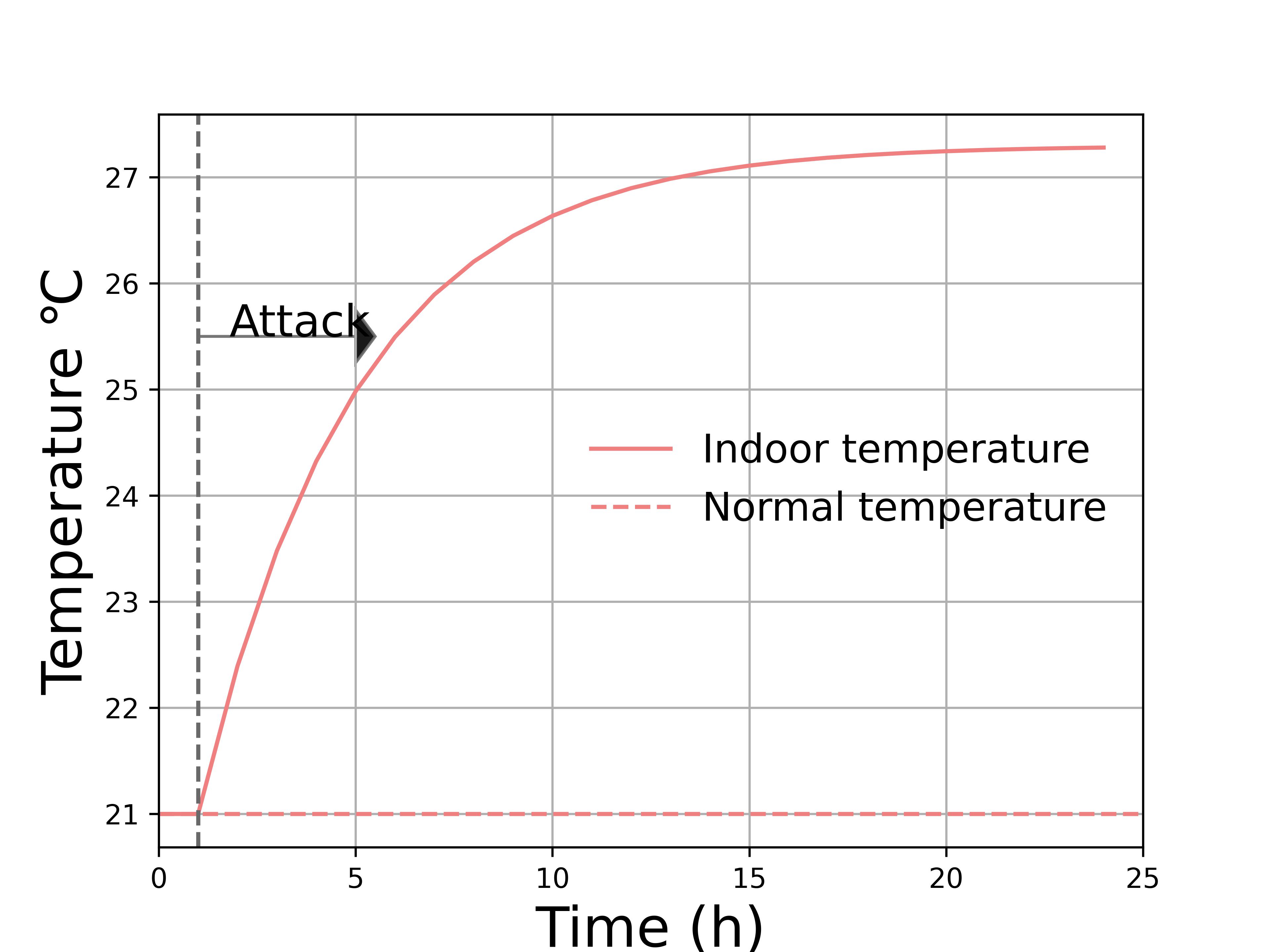}
\caption{Indoor temperature.}
\label{temperature_attack}
\end{minipage}%
\hfill
\begin{minipage}[t]{0.48\linewidth}
\centering
\includegraphics[height=3.3cm,width=4.2cm]{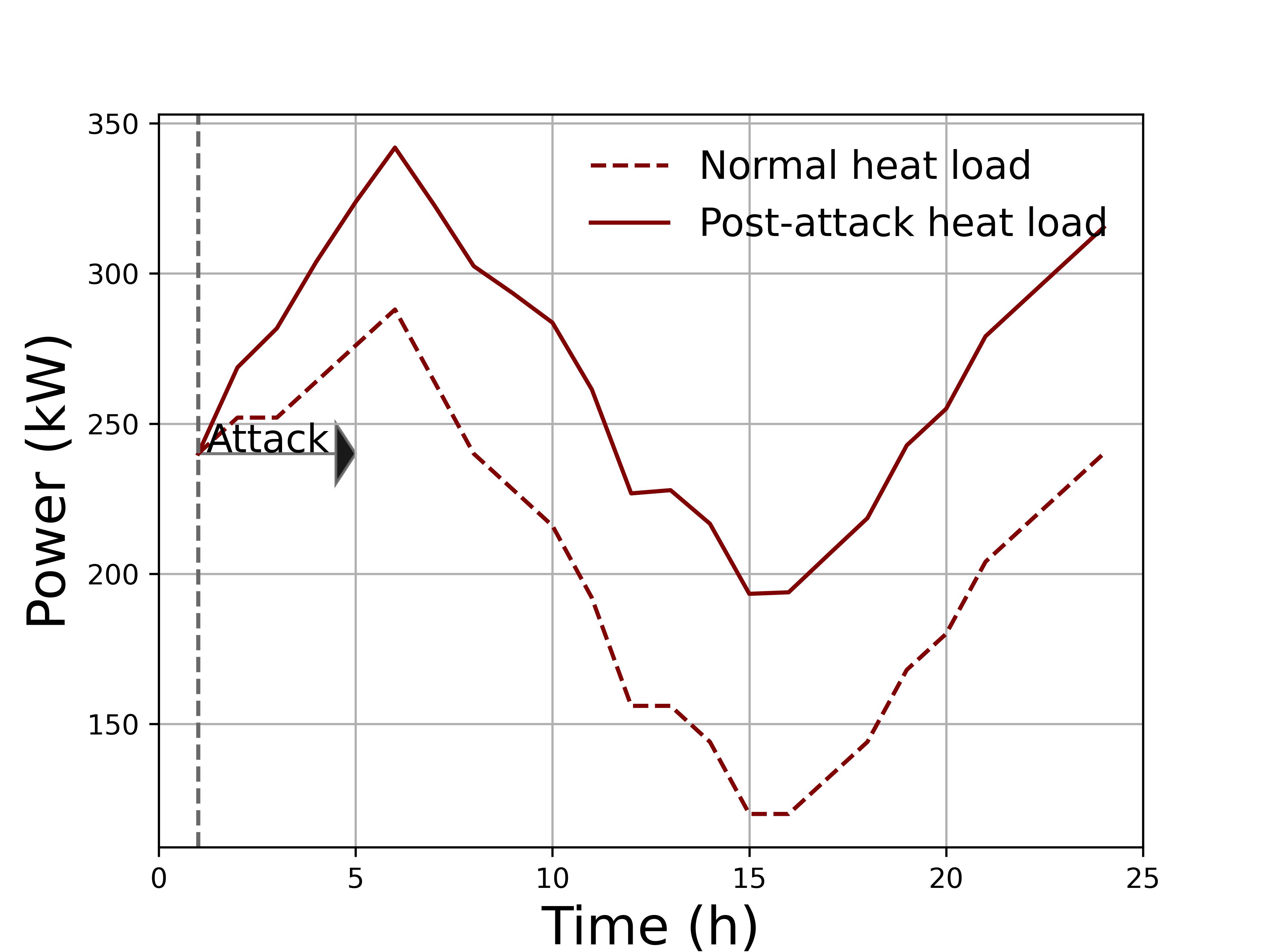}
\caption{Heat load.}
\label{heat_load_attack}
\end{minipage}
\end{figure}

In this study we use HLR attack as described in Eq. (\ref{Eq25}). The change in indoor temperature and the heat load demand after the attack are shown in Figs. \ref{temperature_attack} and \ref{heat_load_attack}.

As shown in Fig. \ref{temperature_attack}, after the implementation of ITDSA, the indoor temperature gradually rises. As shown in Fig. \ref{heat_load_attack}, the increasing indoor temperature causes a gradual increase in heat load demand.

\vspace{-3mm}

To evaluate the impact of cyber-attacks on untrained decision networks and the resilience of our approach, we compared scheduling in four modes:

Mode 1: SAC-based IES scheduling without cyber-attack.

Mode 2: SAC-based IES scheduling with cyber-attack.

Mode 3: Proposed method for IES scheduling without cyber-attack.

Mode 4: Proposed method for IES scheduling with cyber-attack.

\subsubsection{Scheduling Strategy Analysis under Attack Conditions}
Comparing the energy supply of each energy subsystem in Modes 1 and 2, the results are shown in Figs. \ref{mode1_2_ele}-\ref{mode1_2_gas}.

Note from the following figures that there the energy scheduling strategies of Modes 1 and 2 are clearly different. This suggests that decision networks that have not been trained for defense are often highly vulnerable. While they may be able to obtain optimal scheduling policies under normal conditions, they can experience drastic changes in scheduling policies after being subjected to cyber-attacks on the system state. This can lead to a significant decrease in scheduling quality.

\begin{figure}[htbp]
\centering  
\subfigure[]{   
\begin{minipage}[t]{0.49\linewidth}\centering    
\includegraphics[height=3.3cm,width=4.2cm]{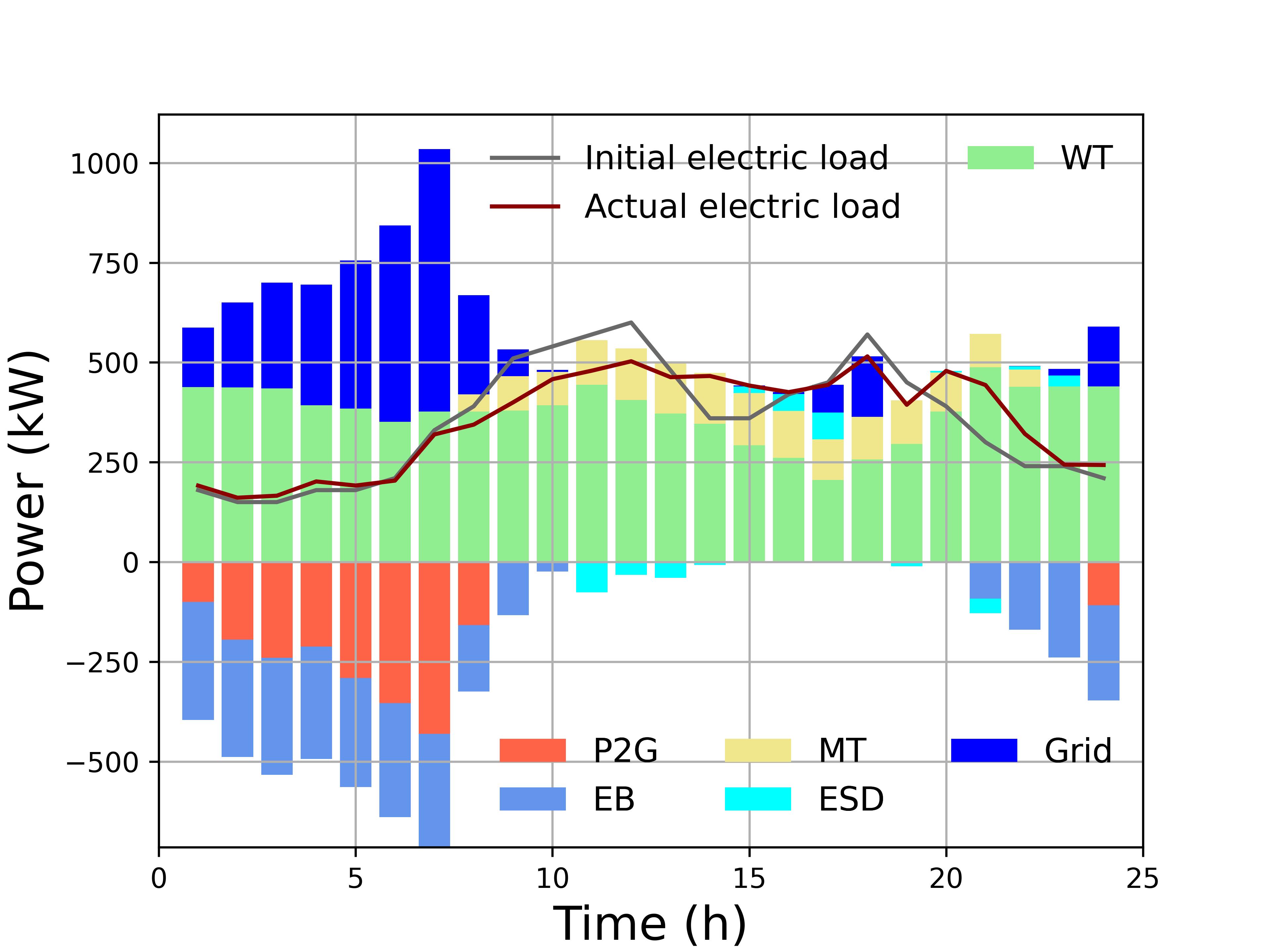}\end{minipage}}\subfigure[]{ 
\begin{minipage}[t]{0.49\linewidth}
\centering    
\includegraphics[height=3.3cm,width=4.2cm]{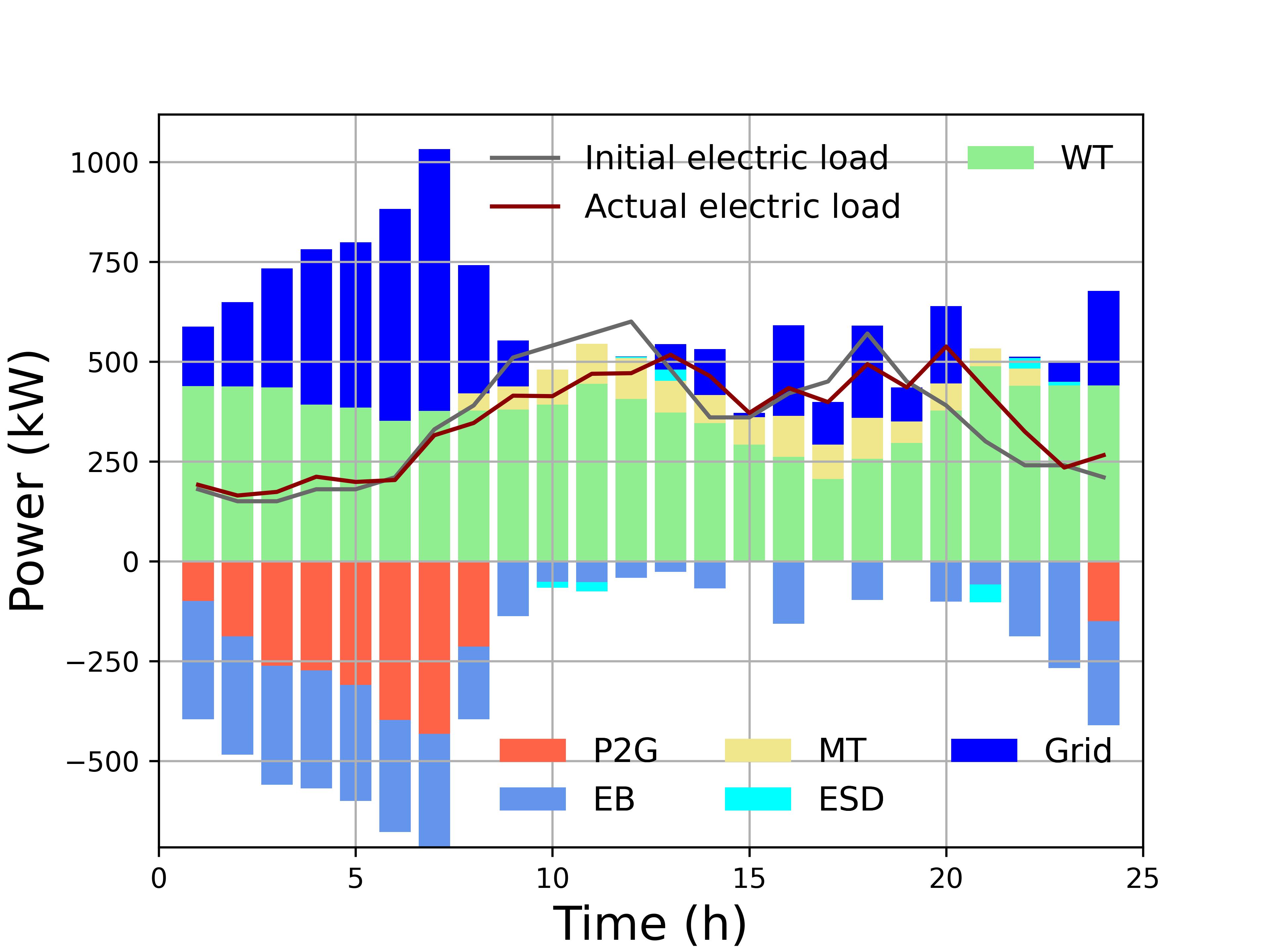}
\end{minipage}
}
\caption{Energy scheduling strategy in Modes 1(a) and 2(b) for the electrical subsystem.}    
\label{mode1_2_ele}    
\end{figure}

\begin{figure}[htbp]
\centering  
\subfigure[]{   
\begin{minipage}[t]{0.49\linewidth}\centering    
\includegraphics[height=3.3cm,width=4.2cm]{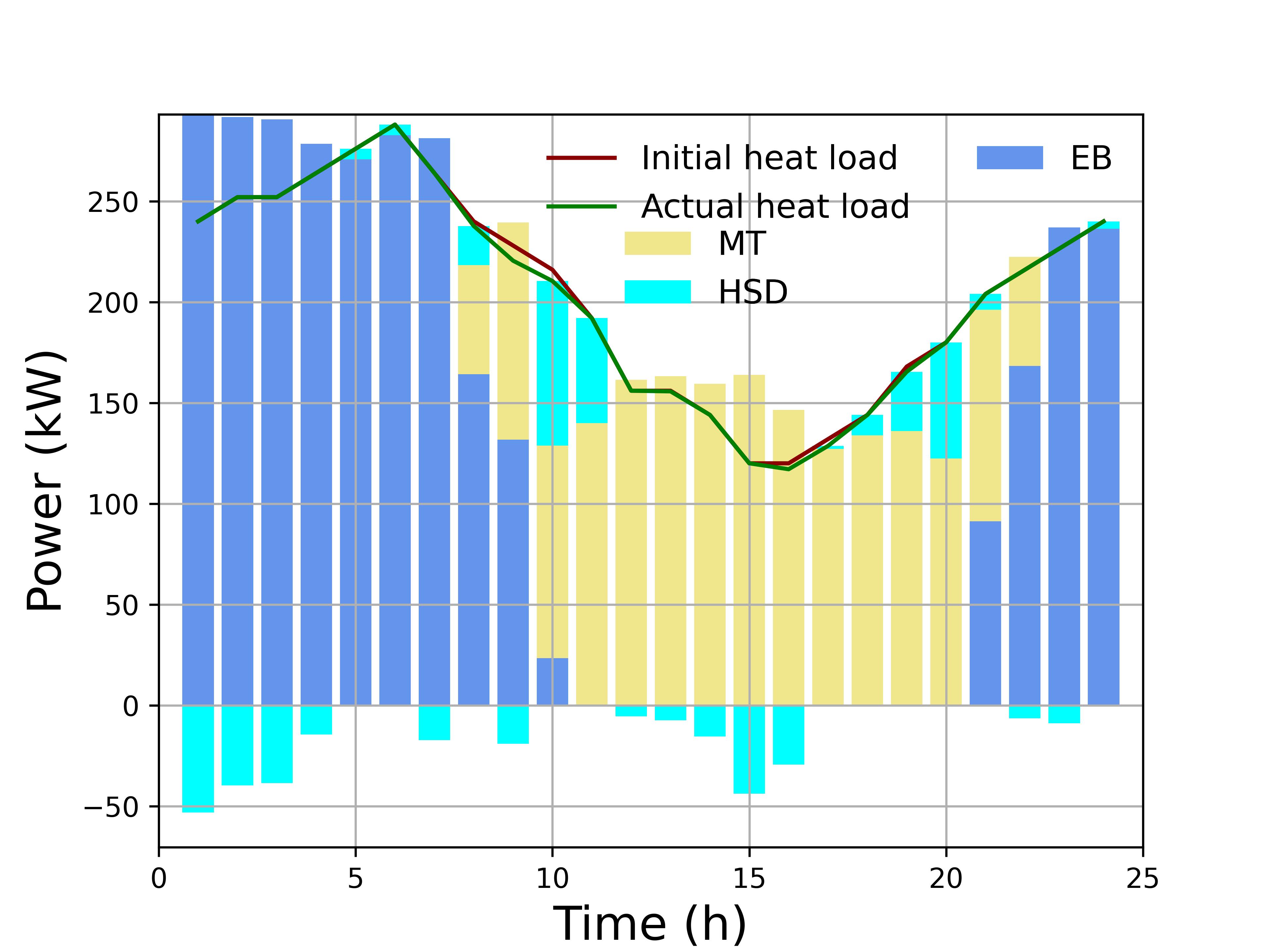}\end{minipage}}\subfigure[]{ 
\begin{minipage}[t]{0.49\linewidth}
\centering    
\includegraphics[height=3.3cm,width=4.2cm]{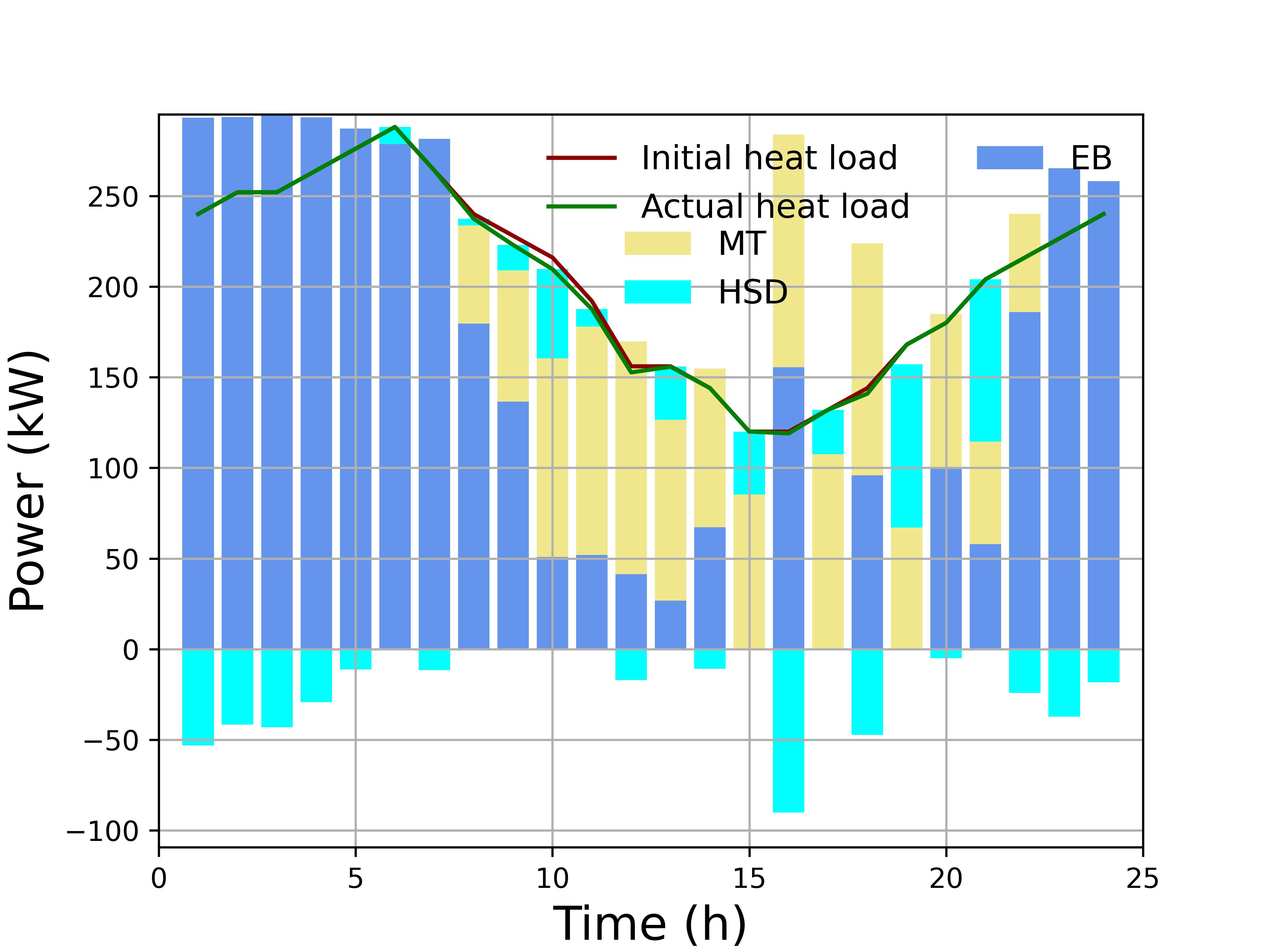}
\end{minipage}
}
\caption{Energy scheduling strategy in Modes 1(a) and 2(b) for heating subsystem.}    
\label{mode1_2_heat}    
\end{figure}

\begin{figure}[htbp]
\centering  
\subfigure[]{   
\begin{minipage}[t]{0.49\linewidth}\centering    
\includegraphics[height=3.3cm,width=4.2cm]{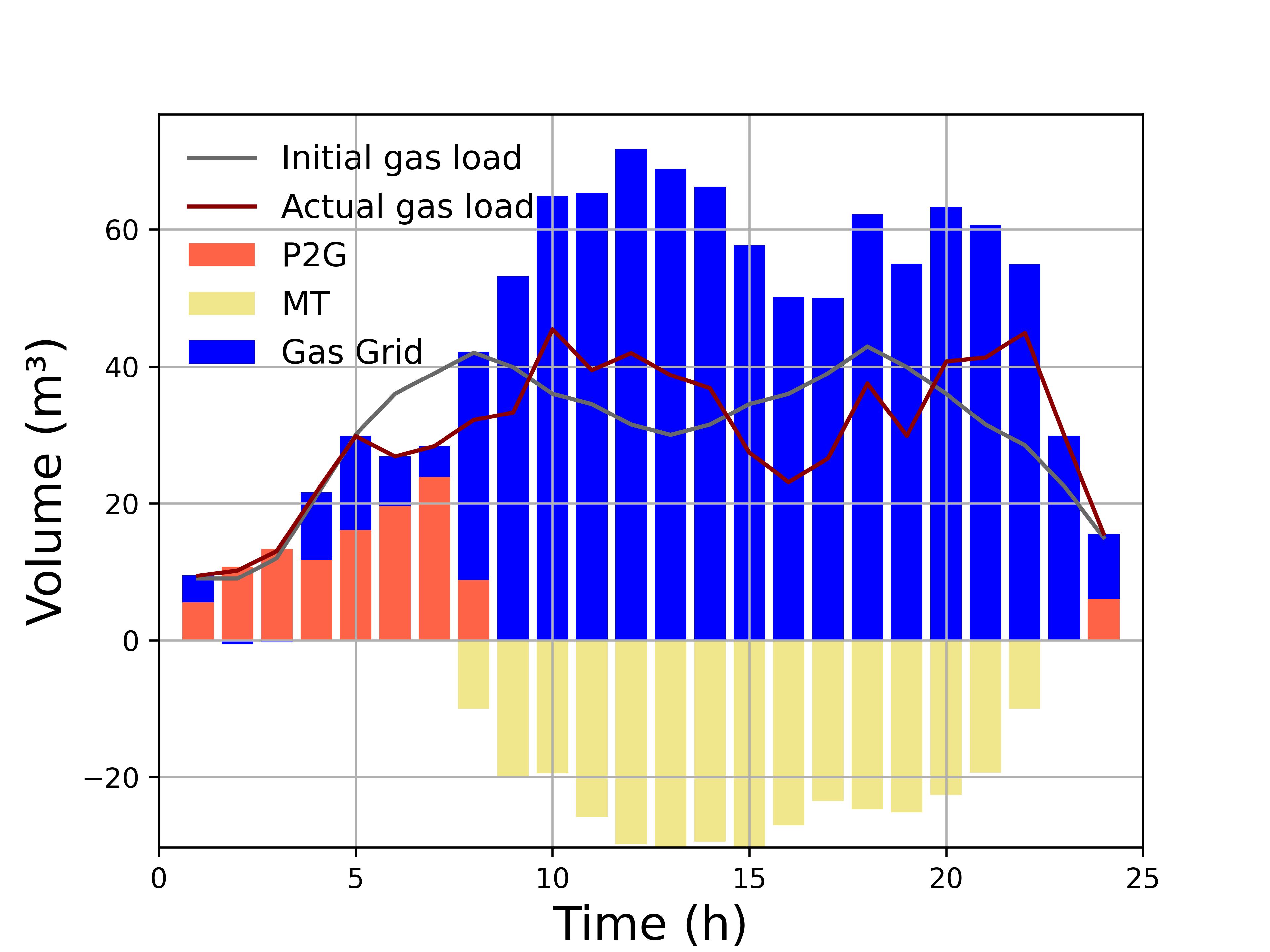}\end{minipage}}\subfigure[]{ 
\begin{minipage}[t]{0.49\linewidth}
\centering    
\includegraphics[height=3.3cm,width=4.2cm]{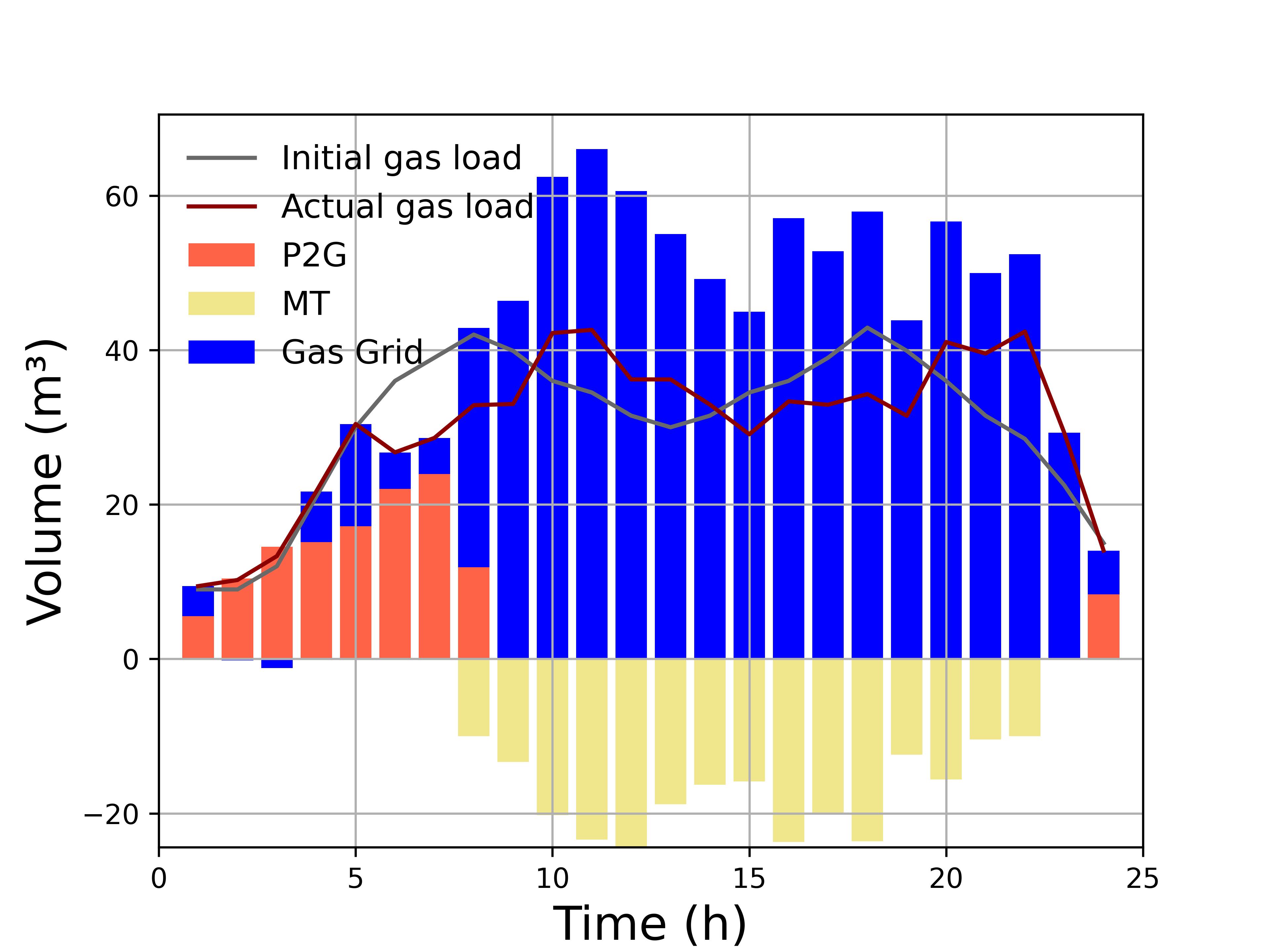}
\end{minipage}
}
\caption{Energy scheduling strategy in Modes 1(a) and 2(b) for natural gas subsystem.}    
\label{mode1_2_gas}    
\end{figure}

Fig. \ref{mode1_2_ele} shows that HLR attacks significantly increase EB output due to false heat load data. Fig. \ref{mode1_2_heat} illustrates that the exaggerated heat demand, arising from manipulated heat load data, leads to a marked escalation in the engagement of HSD. This proactive involvement is aimed at counterbalancing the effects of the spurious demand. Such a trend highlights the potential of thermal storage reserves in providing a degree of resilience against the impacts of the attack. Fig. \ref{mode1_2_gas} indicates MT output reduction due to EB's excessive output, with EB and HSD handling most heat load. 

\subsubsection{Economic Analysis}
To demonstrate the effectiveness of our method, four operational scenarios were compared, differing in HSD participation and IDR application. Scenarios 3 and 4 exclude HSD, while Scenarios 2 and 3 operate without IDR. Table \ref{tab4} shows the IES net profit for each scenario across four scenarios:

Scenario 1: Includes HSD and applies IDR.

Scenario 2: Includes HSD but doesn't apply IDR.

Scenario 3: Excludes HSD but applies IDR.

Scenario 4: Excludes HSD and doesn't apply IDR.

\begin{table}[ht]
    \centering
    \caption{Comparison of economy for different operating Scenarios}\label{tab4}
        \begin{tabular}{ccccc}
        \specialrule{0.5pt}{0.5pt}{0.5pt}
        \specialrule{0.5pt}{0.5pt}{0.5pt}
        \multirow{2}{*}{Operating Scenarios} & \multicolumn{4}{c}{Net profit of IES(¥)} \\
        \cmidrule{2-5}
                                             & Mode 1  & Mode 2  & Mode 3  & Mode 4  \\
        \midrule
        Scenario 1                           & 8746.15 & 8147.27 & 8982.55 & 8934.95 \\
        Scenario 2                           & 7912.12 & 7512.22 & 7933.98 & 7933.14 \\
        Scenario 3                           & 7741.77 & 6796.90 & 8443.16 & 8437.37 \\
        Scenario 4                           & 7067.60 & 6351.33 & 7378.01 & 7377.91 \\
        \specialrule{0.5pt}{0.5pt}{0.5pt}
        \specialrule{0.5pt}{0.5pt}{0.5pt}
        \end{tabular}
    \label{tab:addlabel4}%
\end{table}

Table \ref{tab4} reveals that our method yields higher profits across various scenarios, demonstrating resilience and economic efficiency in normal and cyber-attack conditions. In normal operation, profits increase significantly with the implementation of the IDR program, indicating effective user energy usage guidance. Under cyber-attack, our method's defensive training ensures less volatility and increased profits, showcasing the resilience of our decision-making network. Additionally, the involvement of HSD enhances IES's economic efficiency and resilience, mitigating attack damages by consuming excess heat load, as shown by notable profit increases with the attacks. The improvement in economic indicators demonstrates the resilience capabilities conferred by our proposed method. Profit calculations encompass not only the direct costs related to energy production and consumption but also potential penalties and losses that may be incurred in scenarios of network attacks.

\vspace{-2mm}
\section{Conclusion}
A resilient scheduling method based on DRL for IES effectively addresses multiple uncertainties from RES, loads, and cyber-attacks, ensuring stable supply and enhanced utilization of renewable energy sources. Simulations in North China yielded these conclusions:

(1) The method effectively manages source load uncertainties and resists cyber-attacks, ensuring stable energy supply. It increases renewable energy use via P2G and EB, with a pricing mechanism that guides user DR, particularly during cyber-attacks, and HSD devices that mitigate HLR attack damages.

(2) The SA-MDP model accurately represents scheduling challenges during cyber-attacks. The SA-SAC algorithm, with an action space regularizer, surpasses the SAC in resilience and performance, both under attack and in normal conditions.

(3) Test results show the algorithm's success in creating a resilient IES scheduling strategy with superior economic performance, handling multi-uncertainties and suitable for real-time scheduling across various scenarios.

In future work, blockchain technology will be used to enhance data verification and secure scheduling commands. Specifically, blockchain-enabled smart contracts will automate demand response and energy trading, reducing manual intervention and improving system efficiency and transparency. Additionally, combining adaptive scheduling, deep reinforcement learning, and blockchain’s immutable records is expected to greatly enhance the system's resilience and reliability against evolving network threats.

\bibliographystyle{IEEEtran}
\bibliography{Attack-Resilient.bbl}

\begin{thebibliography}{10}
\providecommand{\url}[1]{#1}
\csname url@samestyle\endcsname
\providecommand{\newblock}{\relax}
\providecommand{\bibinfo}[2]{#2}
\providecommand{\BIBentrySTDinterwordspacing}{\spaceskip=0pt\relax}
\providecommand{\BIBentryALTinterwordstretchfactor}{4}
\providecommand{\BIBentryALTinterwordspacing}{\spaceskip=\fontdimen2\font plus
\BIBentryALTinterwordstretchfactor\fontdimen3\font minus \fontdimen4\font\relax}
\providecommand{\BIBforeignlanguage}[2]{{%
\expandafter\ifx\csname l@#1\endcsname\relax
\typeout{** WARNING: IEEEtran.bst: No hyphenation pattern has been}%
\typeout{** loaded for the language `#1'. Using the pattern for}%
\typeout{** the default language instead.}%
\else
\language=\csname l@#1\endcsname
\fi
#2}}
\providecommand{\BIBdecl}{\relax}
\BIBdecl

\bibitem{wang2018review}
D.~Wang, L.~Liu, H.~Jia, W.~Wang, Y.~Zhi, Z.~Meng, and B.~Zhou, ``Review of key problems related to integrated energy distribution systems,'' \emph{CSEE Journal of Power and Energy Systems}, vol.~4, no.~2, pp. 130--145, 2018.

\bibitem{liu2020region}
L.~Liu, D.~Wang, K.~Hou, H.~Jia, and S.~Li, ``Region model and application of regional integrated energy system security analysis,'' \emph{Applied Energy}, vol. 260, p. 114268, 2020.

\bibitem{lu2020hydraulic}
S.~Lu, W.~Gu, C.~Zhang, K.~Meng, and Z.~Dong, ``Hydraulic-thermal cooperative optimization of integrated energy systems: A convex optimization approach,'' \emph{IEEE Transactions on Smart Grid}, vol.~11, no.~6, pp. 4818--4832, 2020.

\bibitem{che2018mitigating}
L.~Che, X.~Liu, and Z.~Li, ``Mitigating false data attacks induced overloads using a corrective dispatch scheme,'' \emph{IEEE Transactions on Smart Grid}, vol.~10, no.~3, pp. 3081--3091, 2018.

\bibitem{liang20162015}
G.~Liang, S.~R. Weller, J.~Zhao, F.~Luo, and Z.~Y. Dong, ``The 2015 ukraine blackout: Implications for false data injection attacks,'' \emph{IEEE Transactions on Power Systems}, vol.~32, no.~4, pp. 3317--3318, 2016.

\bibitem{zheng2019stochastic}
J.~Zheng, Y.~Kou, M.~Li, and Q.~Wu, ``Stochastic optimization of cost-risk for integrated energy system considering wind and solar power correlated,'' \emph{Journal of Modern Power Systems and Clean Energy}, vol.~7, no.~6, pp. 1472--1483, 2019.

\bibitem{emrani2021optimal}
P.~Emrani-Rahaghi and H.~Hashemi-Dezaki, ``Optimal scenario-based operation and scheduling of residential energy hubs including plug-in hybrid electric vehicle and heat storage system considering the uncertainties of electricity price and renewable distributed generations,'' \emph{Journal of Energy Storage}, vol.~33, p. 102038, 2021.

\bibitem{li2021optimal}
Y.~Li, B.~Wang, Z.~Yang, J.~Li, and G.~Li, ``Optimal scheduling of integrated demand response-enabled community-integrated energy systems in uncertain environments,'' \emph{IEEE Transactions on Industry Applications}, vol.~58, no.~2, pp. 2640--2651, 2021.

\bibitem{zhao2021coordinated}
B.~Zhao, A.~J. Lamadrid, R.~S. Blum, and S.~Kishore, ``A coordinated scheme of electricity-gas systems and impacts of a gas system fdi attacks on electricity system,'' \emph{International Journal of Electrical Power \& Energy Systems}, vol. 131, p. 107060, 2021.

\bibitem{shayan2019network}
H.~Shayan and T.~Amraee, ``Network constrained unit commitment under cyber attacks driven overloads,'' \emph{IEEE Transactions on Smart Grid}, vol.~10, no.~6, pp. 6449--6460, 2019.

\bibitem{liu2016false}
X.~Liu and Z.~Li, ``False data attacks against ac state estimation with incomplete network information,'' \emph{IEEE Transactions on Smart Grid}, vol.~8, no.~5, pp. 2239--2248, 2016.

\bibitem{zhao2021cyber}
P.~Zhao, Z.~Cao, D.~D. Zeng, C.~Gu, Z.~Wang, Y.~Xiang, M.~Qadrdan, X.~Chen, X.~Yan, and S.~Li, ``Cyber-resilient multi-energy management for complex systems,'' \emph{IEEE Transactions on Industrial Informatics}, vol.~18, no.~3, pp. 2144--2159, 2021.

\bibitem{ding2022cyber}
S.~Ding, W.~Gu, S.~Lu, R.~Yu, and L.~Sheng, ``Cyber-attack against heating system in integrated energy systems: Model and propagation mechanism,'' \emph{Applied Energy}, vol. 311, p. 118650, 2022.

\bibitem{li2023optimal}
Y.~Li, F.~Bu, Y.~Li, and C.~Long, ``Optimal scheduling of island integrated energy systems considering multi-uncertainties and hydrothermal simultaneous transmission: A deep reinforcement learning approach,'' \emph{Applied Energy}, vol. 333, p. 120540, 2023.

\bibitem{zhang2021soft}
B.~Zhang, W.~Hu, D.~Cao, T.~Li, Z.~Zhang, Z.~Chen, and F.~Blaabjerg, ``Soft actor-critic--based multi-objective optimized energy conversion and management strategy for integrated energy systems with renewable energy,'' \emph{Energy Conversion and Management}, vol. 243, p. 114381, 2021.

\bibitem{li2023deep}
Y.~Li, C.~Yu, M.~Shahidehpour, T.~Yang, Z.~Zeng, and T.~Chai, ``Deep reinforcement learning for smart grid operations: algorithms, applications, and prospects,'' \emph{Proceedings of the IEEE}, vol. 111, no.~9, pp. 1055--1096, 2023.

\bibitem{tajalli2020resilient}
S.~Z. Tajalli, M.~Mardaneh, E.~Taherian-Fard, A.~Izadian, A.~Kavousi-Fard, M.~Dabbaghjamanesh, and T.~Niknam, ``Dos-resilient distributed optimal scheduling in a fog supporting iiot-based smart microgrid,'' \emph{IEEE Transactions on Industry Applications}, vol.~56, no.~3, pp. 2968--2977, 2020.

\bibitem{chu2022mitigating}
Z.~Chu, S.~Lakshminarayana, B.~Chaudhuri, and F.~Teng, ``Mitigating load-altering attacks against power grids using cyber-resilient economic dispatch,'' \emph{IEEE Transactions on Smart Grid}, 2022.

\bibitem{huang2021distributed}
B.~Huang, Y.~Li, F.~Zhan, Q.~Sun, and H.~Zhang, ``A distributed robust economic dispatch strategy for integrated energy system considering cyber-attacks,'' \emph{IEEE Transactions on Industrial Informatics}, vol.~18, no.~2, pp. 880--890, 2021.

\bibitem{zhao2020cyber}
P.~Zhao, C.~Gu, Z.~Cao, D.~Xie, F.~Teng, J.~Li, X.~Chen, C.~Wu, D.~Yu, X.~Xu \emph{et~al.}, ``A cyber-secured operation for water-energy nexus,'' \emph{IEEE Transactions on Power Systems}, vol.~36, no.~4, pp. 3105--3117, 2020.

\bibitem{hua2019optimal}
H.~Hua, Y.~Qin, C.~Hao, and J.~Cao, ``Optimal energy management strategies for energy internet via deep reinforcement learning approach,'' \emph{Applied Energy}, vol. 239, pp. 598--609, 2019.

\bibitem{nakabi2021deep}
T.~A. Nakabi and P.~Toivanen, ``Deep reinforcement learning for energy management in a microgrid with flexible demand,'' \emph{Sustainable Energy, Grids and Networks}, vol.~25, p. 100413, 2021.

\bibitem{mao2019pmv}
N.~Mao, J.~Hao, T.~He, M.~Song, Y.~Xu, and S.~Deng, ``Pmv-based dynamic optimization of energy consumption for a residential task/ambient air conditioning system in different climate zones,'' \emph{Renewable Energy}, vol. 142, pp. 41--54, 2019.

\bibitem{ku2014automatic}
K.~Ku, J.~Liaw, M.~Tsai, and T.~Liu, ``Automatic control system for thermal comfort based on predicted mean vote and energy saving,'' \emph{IEEE Transactions on Automation Science and Engineering}, vol.~12, no.~1, pp. 378--383, 2014.

\bibitem{li2021coordinating}
Y.~Li, M.~Han, Z.~Yang, and G.~Li, ``Coordinating flexible demand response and renewable uncertainties for scheduling of community integrated energy systems with an electric vehicle charging station: A bi-level approach,'' \emph{IEEE Transactions on Sustainable Energy}, vol.~12, no.~4, pp. 2321--2331, 2021.

\bibitem{zang2019robust}
H.~Zang, M.~Geng, M.~Xue, X.~Mao, M.~Huang, S.~Chen, Z.~Wei, and G.~Sun, ``A robust state estimator for integrated electrical and heating networks,'' \emph{IEEE Access}, vol.~7, pp. 109\,990--110\,001, 2019.

\bibitem{zhang2020robust}
H.~Zhang, H.~Chen, C.~Xiao, B.~Li, M.~Liu, D.~Boning, and C.-J. Hsieh, ``Robust deep reinforcement learning against adversarial perturbations on state observations,'' \emph{Advances in Neural Information Processing Systems}, vol.~33, pp. 21\,024--21\,037, 2020.

\bibitem{zhang2019towards}
H.~Zhang, H.~Chen, C.~Xiao, S.~Gowal, R.~Stanforth, B.~Li, D.~Boning, and C.-J. Hsieh, ``Towards stable and efficient training of verifiably robust neural networks,'' \emph{arXiv preprint arXiv:1906.06316}, 2019.

\end{thebibliography}

\end{document}